\documentclass{article}
\usepackage{amssymb,cite,multirow} 
\def\1ad{\mbox{\normalsize $^1$}}
\def\2ad{\mbox{\normalsize $^2$}}
\def\3ad{\mbox{\normalsize $^3$}}
\def\4ad{\mbox{\normalsize $^4$}}
\def\5ad{\mbox{\normalsize $^5$}}
\def\6ad{\mbox{\normalsize $^6$}}
\def\7ad{\mbox{\normalsize $^7$}}
\def\8ad{\mbox{\normalsize $^8$}}

\makeatletter
\@addtoreset{equation}{section}
\makeatother
\renewcommand{\theequation}{\thesection.\arabic{equation}}
 
\def\beq{\begin{equation}}                     %
\def\eeq{\end{equation}}                       %
\def\bea{\begin{eqnarray}}                     
\def\eea{\end{eqnarray}}                       
\def\nn{\nonumber} 
\def\de {\partial} 
\def\cy {Calabi--Yau}

\def\clss {Clifford(6,6)} 
 
\def\0 {\nonumber} 
\def\del{\partial} 
\def\bi{\bar \imath}
\def\Ox{\Omega} 
\def\bj{\bar \jmath}
\def\minicent#1#2{
  \begin{minipage}{#1 cm}
    \begin{center}
     #2 
    \end{center}
  \end{minipage}
}
\newcommand{\sla}{\slash\!\!\!}
\def\sei{e^{i J}\!\!\!\!\! 
\begin{picture}(10,10)
\put(0,0){\line(1,2){5}}
\end{picture}
}
\def\seim{e^{-i J}\!\!\!\!\!\! 
\begin{picture}(10,10)
\put(0,0){\line(1,2){5}}
\end{picture}
}
              
\begin{document}
\setcounter{page}{0}
\begin{titlepage}
\titlepage
\rightline{hep-th/0406137}
\rightline{CPHT-RR 019.0504}
\rightline{LPTENS-04/32}
\vskip 3cm
\centerline{{ \bf \Large Supersymmetric Backgrounds}}
\vskip .5cm
\centerline{{ \bf \Large from}}
\vskip .5cm
\centerline{{ \bf \Large Generalized Calabi-Yau Manifolds}}
\vskip 1.5cm
\centerline{Mariana Gra{\~n}a$^{a,b}$, Ruben
Minasian$^a$, Michela Petrini$^{a,c}$
and Alessandro Tomasiello$^a$}
\begin{center}
\em $^a$Centre de Physique Th{\'e}orique, Ecole
Polytechnique
\\91128 Palaiseau Cedex, France\\
\vskip .4cm
$^b$Laboratoire de Physique Th{\'e}orique, Ecole Normale
Sup{\'e}rieure\\
24, Rue Lhomond 75231 Paris Cedex 05, France\\
\vskip .4cm
$^c$Laboratoire de Math{\'e}matiques et Physique Th{\'e}orique,
 Universit{\'e} Fran{\c c}ois Rabelais\\ 
Parc de Grandmont 37200, Tours, France

\end{center}
\vskip 1.5cm  
\begin{abstract}

We show that  the supersymmetry transformations for type II string theories on
six-manifolds can be written as differential conditions on a pair of pure
spinors,  the exponentiated K{\"a}hler form $e^{iJ}$ and the holomorphic form
$\Omega$. The equations are explicitly symmetric under exchange of the
two pure spinors and a choice of even or odd-rank RR field. 
This is mirror symmetry for
manifolds with torsion. Moreover, RR fluxes  affect only one of the two
equations: $e^{iJ}$ is closed under the action of the twisted exterior
derivative in IIA theory, and similarly  $\Omega$ is closed in IIB. Modulo a
different action of the $B$--field, this means that supersymmetric 
SU(3)--structure manifolds are all generalized Calabi-Yau manifolds, as
defined by Hitchin. 
An equivalent, and somewhat more conventional, 
description is given as a set of relations between the components of 
intrinsic torsions modified by the NS flux and the Clifford products of RR 
fluxes with pure spinors, allowing for a 
classification of type II supersymmetric vacua on six-manifolds. We find 
in particular that supersymmetric six-manifolds are always complex for IIB 
backgrounds while they are twisted symplectic for IIA.

\end{abstract}

\vfill
\begin{flushleft}
{\today}\\
\end{flushleft}
\end{titlepage}

\newpage
\large
\section{Introduction}

Compactifications with fluxes have received much attention recently due 
to a number of interesting features. In many ways these can be seen as 
extensions of the more conventional compactifications on Ricci-flat 
manifolds. On the other hand, many aspects of the latter, most 
notably in the case of Calabi-Yau manifolds, still have to find their 
generalized counterparts.  
Mirror symmetry has been one of the most prominent and useful 
features of Calabi-Yau compactifications, and the question of its 
extension to compactifications with fluxes is both of conceptual and 
of practical interest. 

The issue of extending mirror symmetry to compactifications with fluxes 
has been studied recently in \cite{kstt,vafa,glmw,fmt,bdkt}. 
A first question is of course
within which class of manifolds this symmetry should be defined. A natural
proposal comes from the formalism of G--structures, recently used in many
contexts of compactifications with fluxes.
As shown in \cite{glmw,fmt}, 
mirror symmetry can be defined on manifolds of SU(3)
structure, thus generalizing the usual Calabi--Yau case. One of the points
which makes this symmetry non--trivial is that, as expected, 
geometry and NS flux mix in the transformation. On the contrary, RR 
fluxes are mapped 
by mirror symmetry into RR fluxes and  their transformation is
well-understood. However, for many reasons it would be better to have a 
formalism
that would incorporate geometrical data and fluxes in 
a natural way.
This paper is a step in that direction. We will propose to use {\it pure
  spinors} as a formalism to describe SU(3)--structure compactifications. 

The first reason to look for a unifying formalism is essentially checking the
proposal for mirror symmetry given in \cite{fmt}. 
In that paper, a quantitative rule for obtaining
mirror-symmetric backgrounds is given based on the action of the 
twisted covariant
derivative on spinors. From such rule one can read off the exchange of the 
components of the NS flux with the quantities describing the failure of the
integrability of the complex structure and the K{\"a}hler form. As we will
review in section 2, it works
essentially exchanging representations $ 8+1 \leftrightarrow 6+\bar 3$:
$$
(\nabla J + H)_{ijk} \longleftrightarrow (\nabla J - H)_{i\bj\bar k} .
$$
 Though checked
on a number of examples, the formula is
conjectural for the following reason:
it was derived assuming that the manifold and the fluxes  under 
consideration admit 
three Killing 
vectors, and then performing simultaneous T--duality along
the three  isometries. The same procedure is known in the context of
Calabi--Yau mirror symmetry as the SYZ \cite{syz} approach. There, however, the
structure of $T^3$ fibration was derived from considerations of moduli spaces
of branes, which are lacking in compactifications 
with fluxes. However, the formula
obtained  under that assumption is clean enough to be conjectured to be valid
when the $T^3$ fibration structure is not present. 
 
Inclusion of RR fluxes gives a check of the conjecture in the following
sense. Mirror symmetric compactifications should yield the same physics in
four dimensions. In particular, a compactification which preserves
supersymmetry should be sent to another one with the same property.
Since on supersymmetric
backgrounds the total NS and RR contributions to the 
supersymmetry equations sum up to zero, demanding that
mirror symmetry maps supersymmetric backgrounds to supersymmetric
backgrounds allows to check if the proposed NS transformation is
compatible with the known RR one and if so to lend further support for the whole
picture.

It is easy to see that this check has a chance of working  realizing that
two objects are the same: {\it \clss\ spinors } and {\it bispinors}. These
appear in NS and RR sector respectively. 
The \clss\ spinors  appeared in \cite{fmt} in order to interpret the 
mirror symmetry formulae in a more natural way. As far as we are concerned in
this introduction, \clss\ spinors are simply {\it formal sums of
  forms}. (In analogy with  usual spinors, which are often realized as 
formal sums
of $(0,p)$ forms.) Such a spinor is called 
 {\it pure} if it is annihilated by half of the gamma matrices. A pure spinor
 defines an SU(3,3) structure on the bundle $T + T^*$ on the manifold.  
If the spinor is also closed, the manifold is called by Hitchin \cite{hitchin}
a {\it generalized 
\cy}.\footnote{In \cite{gualtieri}, the same name is used for a
different type of manifold, that has two pure spinors whose 
associated generalized complex structures are commuting and integrable.}  
For a SU(3) structure on $T$, there are two pure spinors, which are 
orthogonal and of unit norm. An SU(3) structure is defined by a 
two--form $J$ and a three--form  
$\Omega$ obeying $J\wedge \Omega=0$ and $i\Omega\wedge\bar\Omega = (2J)^3/3!$. 
Then, the two pure spinors are $e^{i J}$ and $\Omega$.
From this point of view it is natural to 
conjecture that mirror symmetry between two SU(3) structure manifolds  
exchanges these two pure spinors. 

It is also possible to incorporate the $B$--field since
multiplying a pure spinor by $e^B$ leaves it pure \cite{hitchin}. 
This is indeed what happened in \cite{fmt}: 
T--duality along $T^3$ (when it is possible) realizes the  exchange  
$$
e^{B+iJ}\longleftrightarrow \ \Omega \ ,
$$
thus motivating the introduction of 
the Hitchin formalism just 
mentioned. In the \cy\ case, this 
exchange is implicit in many applications of mirror symmetry.  

The second fact, that RR fields are described by bispinors, is much more
standard and familiar from the very spectrum of the superstring. In this paper
we will use in many instances that the Clifford(6,6) spinors above can be
identified with bispinors (Clifford(6)$\times$ Clifford(6)) under the
map from forms to elements of the Clifford(6) algebra, $dx^m\to \gamma^m$.

Using this identification, we will be able to show that the
supersymmetry transformations for IIA and IIB can be
written in a unified fashion using formally two pure spinors and a total RR
field of either even or odd rank.  Very schematically
\bea
\label{eq:clif}
\delta \psi_m = [D^H_m+ (\varphi_1 \cdot  F)_m + (\varphi_2 \cdot
F)_{mn}\gamma^n]\epsilon\ ,
\eea
$F$ here is the formal sum of all RR fields, the dot stands for a
Clifford multiplication and $\varphi_1, \varphi_2$ are pure spinors. It is not hard to see that choosing the RR field
$F$ to be even or odd fixes the roles played by each pure spinor which
has to be even or odd as well.  
Mirror symmetry then will
exchange the pure spinors and change the rank of the RR field from odd to even
and vice versa. Essentially formulae as (\ref{eq:clif}) 
come from defining SU(3) structures
in terms of  an {\it ordinary} spinor $\eta$, and then using Fierz 
identities to
express  the two pure spinors as  $e^{i J}= 8 \eta_-\otimes \eta_-^\dagger$,
$\Omega = 8i \eta_+\otimes \eta_-^\dagger$. In these terms, mirror symmetry
can be seen as conjugation on a sector. 

In fact, there is a more precise sense in which the formalism of pure spinors
is relevant. Indeed, in section \ref{sec:pure} we will see  that the 
supersymmetry equations imply
differential equations for the  pure spinors, schematically
\bea
\label{eq:pss}
e^{-f_1} d (e^{f_1}\varphi_1)&=&  H\bullet \varphi_1 \nn \\
e^{-f_2} d (e^{f_2} \varphi_2)&=&  H\bullet \varphi_2 + (F, \varphi) \ .
\eea
The operator $H\bullet$ is a certain action of the $H$ three--form, involving
contractions and wedges but {\it different} from $H\wedge$.
So, in both IIA and IIB there is a ``preferred" pure spinor
(of the same parity as $F$ - namely $e^{iJ}$ in IIA and $\Omega$ in IIB) 
which does
not receive any back reaction from the RR fluxes. 
This property is called (twisted) generalized \cy\ \cite{hitchin}. The twisting
refers to the presence of the $H$ field. In the mathematical literature (and
in some physical applications \cite{rw})
this twisting is actually always appearing in the form $(d + H\wedge)$. It is
interesting to see that in general the inclusion of RR fluxes requires a
different form of twisting than the one usually assumed. Understanding the
origin of this twisting from first principles would be of some importance.

Much of this discussion can be carried out in more conventional (and also 
somewhat more practical from point of view of finding examples) terms.
In supersymmetry 
transformations one can well separate the NS and RR contributions. The 
former are given by components of SU(3) intrinsic torsion modified by 
inclusion of the NS three-form flux. The latter are Clifford products of 
RR fluxes with geometric data (pure spinors again) consistent with 
(\ref{eq:clif}). It turns out that the RR fluxes affect only some of the 
components of the intrinsic torsion (compare to (\ref{eq:pss})), thus 
making the analysis of supersymmetry conditions rather easy and allowing 
for a complete classification of type II theories on six-manifolds. In 
particular we show that the supersymmetric geometries in IIB are always 
complex while in IIA they are twisted symplectic. Mirror symmetry can 
also be seen as well as  the respective mappings of RR-corrected and 
RR-uncorrected sectors in IIA and IIB (in a 
agreement with  $6+\bar 3 \leftrightarrow 8 +1$ rule).

\medskip

This two-level discussion (``spinors" versus ``pure spinors") is reflected in 
the structure of the paper, which has two complementary but self-contained 
parts.  In section \ref{sec:msns} we review the basics of the 
formalism and the way mirror symmetry works
for geometry and $B$--field.  We proceed to describe the general features
of RR supergravity transformations in section \ref{sec:RR}, where we also 
show how these can be put in a 
manifestly mirror symmetric way on manifolds of SU(3) structure.
Analysis of the supersymmetry conditions is presented in section
\ref{sec:susy}. Finally, in section \ref{sec:pure} we discuss these
conditions in terms of pure spinors and show in particular the
correspondence between supersymmetric string vacua and generalized
Calabi-Yau manifolds.

\section{NS flux and geometry}
\label{sec:msns}
In this section we briefly review the action of mirror symmetry on the  NS
sector, to set the stage but also to clarify some points from 
\cite{fmt}.\footnote{The discussion here is far from
being complete and is 
concerned mostly with the spinorial aspects of the formalism. We refer to 
\cite{glmw, fmt} and references therein for a more complete account of 
G-structures and intrinsic torsions.}

We start by briefly introducing the  
notions of SU(3)-structure and intrinsic torsion with the help of which  
we will describe the non-Ricci-flat geometries under consideration.
A manifold with 
SU(3)-structure has all the group-theoretical features of a \cy, 
namely invariant two- and three forms, $J$ and $\Omega$ respectively. 
On a manifold of SU(3) {\it holonomy}, not only $J$ and $\Omega$ 
are well defined, but they are also closed: $dJ=0=d\Omega$. If they are not 
closed, 
$dJ$ and $d\Omega$ give a good measure of how far the manifold is from having 
SU(3) holonomy
\begin{equation}
\label{eq:djdo}
\begin{array}{c}\vspace{.3cm} 
dJ = -\frac{3}{2}\, {\rm Im}(W_1 \bar{\Omega}) + W_4 \wedge J + W_3 \, ,\\
d\Omega = W_1 J^2 + W_2 \wedge J + \bar{W_5} \wedge \Omega\ \, .
\end{array}
\end{equation}
The $W$'s are the $(3 \oplus \bar{3} \oplus 1) \otimes (3 \oplus \bar{3})$
components of the intrinsic torsion: 
$W_1$ is a complex zero--form in $1 \oplus 1$, $W_2$ is a complex primitive 
two--form, so it lies in $8 \oplus 8$, $W_3$ is a real primitive 
$(2,1) \oplus (1,2)$ form and it lies in 
 $6 \oplus \bar{6}$,  $W_4$ is a real one--form in $3 \oplus \bar{3}$, and 
finally $W_5$ is a complex $(1,0)$--form (notice that in (\ref{eq:djdo}) the  
$(0,1)$ part drops out), so its degrees of freedom are again  
$3 \oplus \bar{3}$.  
 
These $W_i$ allow to classify the differential type of any SU(3) structure. 
A simple counting tells 
that there can be a representation of the intrinsic torsion  given simply 
by a six-by-six matrix and a vector. There is indeed an alternative 
definition of the torsions which essentially does this. A SU(3) structure can 
be
defined also by a spinor $\eta$. In terms of this, $J$ and $\Omega$ above are
defined as bilinears: 
$\eta^\dagger \gamma_{mn}\gamma\,\eta = i J_{mn}$ and 
$-i \eta^\dagger \gamma_{mnp}(1+\gamma)\eta = \Omega_{mnp}$, where $\gamma$ is the 
six dimensional chirality operator.

The spinor $\eta$ also gives a basis for all spinors on the manifold: 
$\eta$, $\gamma \eta$ and $\gamma^m \eta$.\footnote{In the following,
differently from notation in \cite{fmt},
we will denote real six--dimensional indices as $m,n,\ldots$ and
holomorphic/antiholomorphic indices as $i,j,\ldots$ 
($\bar i, \bar j,\ldots$).} Anything else in the Clifford
algebra acting on  $\eta$, say $\gamma^{m_1\ldots m_n}$, can be
re-expressed in terms of 
this basis. So, in general we can write \footnote{The notation for $q_m$ and 
$\tilde{q}_m$ is the opposite
as that used in \cite{fmt}, since shortly we will set $\tilde{q}$ to zero by 
normalizing the spinor and keep $q_m$ nonzero.} 
\begin{equation}
\label{eq:deps}
D_m \eta = \left(\tilde q_m + i q_m \gamma + iq_{mn}\gamma^n\right)
\eta\ .
\end{equation}
The $q$'s, defined by this equation, 
are real, and provide just 
another definition of the intrinsic torsion. 
It is immediate to notice that there
is a certain redundancy  in (\ref{eq:deps}): it has
three vectors ($q_m$, $\tilde q_m$ and one from $q_{mn}$), and this  
constrained trio is the counterpart of the more economical pair given by $W_4$
and $W_5$. There are two natural ways of resolving this ambiguity. One, which
was used in \cite{fmt}, consists in noticing that only the $(1,0)$ part of 
$\tilde q_m + i q_m$ appears in $W_4,W_5$, and consequently assuming it has no
$(0,1)$ part. Here we will use another method. Indeed, normalizing
the spinor $\eta$ to have {\it constant} norm allows to set $\tilde q_m=0$.

Having fixed the ambiguity, $q$ and $W$ are simply related by a change of 
basis, which in holomorphic/antiholomorphic indices reads
\begin{equation} 
  \label{eq:qmnhol} 
  \begin{array}{c}\vspace{.3cm}
  q_{ij} = -\frac{i}{8} W^3_{ij} - \frac18\Omega_{ijk}\bar W_4^k \, ,
  \\ 
  q_{i \bj} = -\frac i4 \bar{W}^2_{i\bj} +\frac14 \bar W_1 g_{i\bar j} \, ,
  \end{array}
\qquad  q_i = \frac i2 (W_5 -W_4)_i
\ . 
\end{equation} 
Here $W_i= W_i^+ -i W_i^-$ as usual in the literature, and we have 
defined $W^3_{mn} = W^3_{mpq} {\Omega^{pq}}_n$ 
.  

The $q$'s can also be related to the covariant derivatives of $J$ and
$\Omega$. Let us define the real and imaginary parts of $\Omega$ as
$\Omega= \psi - i \tilde \psi$. In real indices, then, we have 
\begin{equation}
  \label{eq:nablaJ}
q_{mn} = \frac{i}{8} \nabla_m J_{pq} {\tilde \psi}^{pq}_{\ \ n}\ .  
\end{equation}
Also, one can directly think  in terms of the so--called contorsion 
$\kappa_{mnp}$, defined as $\nabla_m \eta = \kappa_{mnp}\gamma^{np}\eta$:
$$
\kappa_{mnp} = \frac{i}{8} \nabla_m J_{ns}J^s_{\ p} +\frac1{288}(\tilde 
\psi^{rst}
\nabla_m \psi_{rst}) J_{np}
$$

The basis (\ref{eq:djdo}) is usually more popular because easier to analyze. 
For instance,
looking at it one immediately concludes that $W_1=W_2=0$ iff the manifold is
complex.  
Indeed the $(2,2)$ part of $d\Omega$, $W_1 J^2 + W_2 \wedge J$,
would be absent in the case of integrable complex structure. 
 
From other side, the spinorial approach treats $J$ and $\Omega$ more
symmetrically and turns out to be much more T--duality friendly. It is indeed
immediate to notice that there is no natural exchange of the quantities $W$ in 
(\ref{eq:djdo}); it was this fact that prompted the definition of
(\ref{eq:deps}). 



 We are ready now to discuss the T--duality/mirror transformation 
\footnote{The 
transformation rules for the fields as well as the working assumptions   
can be found in \cite{fmt}, here we just quote some relevant results.}. 
For a generic metric (with a nontrivial connection on the
fiber) neither $W$'s nor $q$'s can be invariant under T--duality - they
necessarily mix with the flux
\begin{equation} 
  \label{eq:compH} 
  H = -\frac{3}{2}\, {\rm Im}(H^{(1)} \bar{\Omega}) + H^{(3)} \wedge J 
+ H^{(6)} 
\end{equation} 
where the flux components are labeled according to the representation, 
namely $H^{(1)}$ is the $1 \oplus 1$ complex scalar, $H^{(6)}$ the 
$6 \oplus \bar{6}$ real 3-form and $H^{(3)}$ the $3 \oplus \bar{3}$ real 
1-form. Including the flux will lead 
to a complexification of the components of the intrinsic torsion in matching 
representations; however, nontrivial transformations must mix 
different representations. This is for the following reason. 
The two mirrors have two different SU(3) structures since the two 
SU(3) are differently embedded into Spin(6,6), because the fiber directions 
change from the tangent bundle to the cotangent bundle. 
As a result, representations get actually mixed. Indeed, as we noticed 
already, 
the $W$ in (\ref{eq:djdo}) have no natural ``pairing''. 

There are two ways one goes around this problem. 
One is to go to the base, where one can 
define T--duality invariants, and further decompose the SU(3) 
representations thus allowing them to mix. 
Alternatively, we could consider the sum of the tangent and cotangent bundles 
and take sums of representations. Our final formula for mirror symmetry 
will be doing exactly this. 
 
The first approach is the one  proven in \cite{fmt} using T--duality.
Assuming that the initial six--dimensional manifold has a fibration structure,
one can decompose the torsions under the SO(3) of the base, 
which is smaller of the
total SU(3). $W_2$ and $W_3$ get then split 
as $W_2=w_2^s+w_2^a$ ($8\to 5 \oplus 3$) and $W_3=w_3^s+w_3^t$ ($6\to 
5 \oplus 1$). Using the usual expressions for the metric and the B--field 
in the 
torus--fibered case, we get under T--duality 
\begin{equation} 
  \label{eq:wtrans} 
  \begin{array}{c} 
  W_1 - iH_{(1)}  \longleftrightarrow -\overline{(W_1 - iH_{(1)})}\ , \\[2mm] 
  \bar{w}_2^s  \longleftrightarrow w_3^s - iH^{(6)s} \, , \\[2mm] 
  W_5, \bar{w}_2^a \longleftrightarrow W_4 - iH^{(3)}\ . 
  \end{array}\end{equation} 
 T--duality preserves SO(3) representations, as it should since it does not
 act on the base. More interesting is to notice that a complexification
 occurred naturally between certain components of $W$ and flux $H$.
 
Thus one can add in (\ref{eq:deps})
a dependence 
on $H$ to the covariant derivative  (and as a consequence the intrinsic
torsion) 
\begin{equation} 
  \label{eq:dheps} 
  D_m^H \eta = i \left( Q_m \gamma + 
    Q_{mn}\gamma^n\right)  
\eta\ . 
\end{equation} 
The idea is to  construct $D^H$  in such a way 
as to find good T--duality transformation properties afterwards.  Not 
too  surprisingly, one finds that the best definition is exactly 
the same as that in the supergravity supersymmetry 
transformations: $D_m^H\equiv (D_m+\frac{1}{8} H_{mnp}\gamma^{np})$.  
The ``twisted" components of the intrinsic torsion turn out to be diagonal 
under T--duality: elements of the basis transform picking a sign. 
In holomorphic/antiholomorphic indices,
\begin{equation} 
  \label{eq:QMN} 
  \begin{array}{c}\vspace{.3cm}
  Q_{ij} = -\frac{i}{8} \left(W_3 +i H^{(6)}\right)_{ij} -
\frac18\Omega_{ijk}\left(\bar W_4 + i \bar H^{(3)} \right)^k \, ,
  \\ 
  Q_{i \bj} = -\frac i4 \bar{W}^2_{i\bj} - \frac14 \left(\bar W_1 
+ 3i \bar H^{(1)}\right) g_{i\bar j} \, ,
  \end{array}
\qquad  Q_i = \frac i2 (W_5  - W_4 -i H^{(3)})
\ . 
\end{equation} 
So, adding $H$ as $D\to D^H$ complexifies $W$ as  $W+iH$.
Using (\ref{eq:wtrans}), 
one can verify that this ``complexified'' $Q$'s, when restricted to the base,
transform nicely, essentially picking $\pm$. So, the spinorial basis is more
suited for T--duality than the original $W$ basis. 
 
The last step is now to conjecture a transformation rule which might be valid
also in cases which are {\it not}  $T^3$ fibrations. One guideline is the
transformation rule found above. Another one is that, as mentioned in the
introduction, mirror symmetry sends supersymmetric vacua to
supersymmetric vacua. In a sense, T--duality is induced by an 
exchange of $\epsilon_+$ with  $\epsilon_-$. Since we also have $\gamma^{\bar
  i}\epsilon_-=0$, one arrives at
\begin{equation} 
  \label{eq:prop} 
  Q_{ij} \longleftrightarrow - Q_{i\bj}\ , \qquad Q_i \longleftrightarrow 
  -{\bar Q}_i\ . 
\end{equation} 
This is the other way of getting around the lack of natural pairing of
representations in (\ref{eq:djdo}). Qualitatively we have $6+\bar 3
\leftrightarrow 8 +1$.

We can also express this exchange in a way which maybe is more mnemonical. 
Using
(\ref{eq:nablaJ}) we have 
\begin{equation}
  \label{eq:newmirror}
  \left(\nabla J +H\right) _{ijk} \longleftrightarrow  
\left(\nabla J - H\right)_{i \bj\bar k}\ ;
\end{equation}
notice that $ \nabla_i J_{j\bar k}$ is automatically zero by hermiticity of
$J$ ($J_{i \bj}= i g_{i \bj}$) and the fact that the connection is 
Levi--Civita. In fact, in this form the symmetry $6+\bar 3
\leftrightarrow 8 +1$ was already noticed long ago by Salamon \cite{salamon}
in a different context predating the mirror symmetry.\footnote{We thank
S.~Salamon for pointing out his
work and   for useful discussions on this section.} 
On a manifold $M$ of  any dimension, one can consider the
bundle ${\cal T}$ of almost complex structures, 
or, equivalently, of lines of pure spinors: this is called 
{\it twistor} bundle. 
In \cite{salamon} it was shown how to define two almost complex
structures $F_i$  on the total space of ${\cal T}$. 
There are relations between the 
behaviour of these two almost complex structures $F_i$ 
and our split $q_{ij}$ versus $q_{i \bj}$ above. For example, a section 
of the twistor bundle is a
holomorphic submanifold with respect to the almost complex structure $F_1$ if
$q_{i \bj}=0$, whereas it is holomorphic with respect to $F_2$ if 
$q_{ij}=0$.  
(Note that neither (\ref{eq:newmirror}) nor the 
pair of the complex structures on the twistor bundle refer to $\Omega$ and 
therefore do not involve $W_5$. Thus some of the arguments here may be
extended to manifolds where the 
structure group is given by U(3) rather than SU(3).) These results seem 
clearly to be relevant to a further
understanding of mirror symmetry, and it would be nice to realize them
physically. Maybe a model on the twistor space is the right way to prove
mirror symmetry from first principles. 

We will conclude this section 
with a brief review of the conditions for 
supersymmetry in the case with $H$-flux only \cite{hull,strominger}, which in 
this language become conditions for the vanishing of components of $Q$'s.
In many ways this example sets the stage for our discussion in a sense that it
gives two basic equations for the pure spinors, which then may or may not be
modified by the RR back reaction.

To have supersymmetry it is enough that one chirality, say $\eta_+$, is  
annihilated by $D^H$. From (\ref{eq:deps}) we have 
\begin{equation} 
  \label{eq:susy} 
  \begin{array}{rcl}\vspace{.3cm} 
  D^H_i \eta_+ &=& i  Q_i \eta_+ 
                     + i Q_{ij}\gamma^j\eta_- \, ,
\\ 
  D^H_{\bi} \eta_+ &=& i  Q_{\bi} \eta_+ 
                            + i Q_{\bi j}\gamma^j \eta_- \, .
  \end{array} 
\end{equation} 
Here $\eta$ is also normalized to one; as we said, the spinors preserved
by supersymmetry often do not have this property, but we can always rescale
them. Notice that $Q_{i \bj}$ and $Q_{\bi \bj}$ have disappeared 
from $D^H \eta_+$, 
because $\eta_-$, being a Clifford vacuum, is annihilated by 
$\gamma^{\bi}$.  
From (\ref{eq:susy}) it follows directly that the complexified $Q_{ij}$ and  
$Q_{\bi j}$ have to vanish. These will say that the complexified 
$W_3$ has to be purely antiholomorphic, which in more usual terms means  
of type $(1,2)$ (this is the condition $W_3= *H_3$) and that $W_2$ has to 
vanish. The vectors require a little more care because usually the dilaton  
is rescaled in the metric (as a warping) and in the spinor itself.  
More generally it is clear that one can use the gamma matrices identities 
mentioned above to reduce the expression to a form like (\ref{eq:deps}), and 
then use (\ref{eq:qmnhol}). 
 
In this case, supersymmetry is trivially consistent with the
proposal for mirror symmetry (\ref{eq:prop}), since both $Q_{ij}$ and $Q_{i
  \bj}$ are zero.  In this form the duality 
might seem a little tautological, in the sense that it sends a 
supersymmetric vacuum in another one in an obvious way. Compare 
however with the usual mirror symmetry: a \cy\ is sent to another \cy, 
and the non triviality lies in the exchange of K{\"a}hler and complex 
structure moduli. A better understanding of the moduli in the flux
compactifications is one important open question. 

Another way to see that the conditions for this case are consistent with the 
mirror symmetry proposal is to put them in a manifest symmetric way
\begin{equation} 
  \label{eq:str} 
  (D_m +\frac14 H_{mnp}dx^n \iota^p ) e^{i J} = 0 \ ,\qquad 
  (D_m +\frac14 H_{mnp}dx^n\iota^p ) \Omega = 0 \ , 
\end{equation} 
where $\iota^m\equiv g^{mn}\iota_{\partial_n}$ \footnote{\label{iota} $\iota_{\partial_n}$: $\Lambda^p T^* \rightarrow
\Lambda^{p-1} T^*$, $\iota_{\partial_n} dx^{i_1} \wedge ... \wedge dx^{i_p} = p \delta^{[i_1}_n dx^{i_2} \wedge ... \wedge dx^{i_p]}$.} 
. $D_m$ can be written as a covariant derivative for bispinors or as the usual Levi-Civita covariant derivative on forms. The object $e^{iJ}$ is a 
formal sum of forms; its meaning will be explained in section 
\ref{sec:pure}.
In section \ref{sec:pure} we 
will also see how (\ref{eq:str}) get modified in presence of RR fields.
 
\section{RR fluxes} 
\label{sec:RR}

In this section we consider the introduction of RR fluxes and analyze how this
affects the supersymmetry conditions and mirror symmetry.
The idea is to generalize what done in the previous section for the NS fluxes.
Just as the entire NS contribution to 
the covariant derivative of the invariant spinor was summarized in $Q$'s 
(see (\ref{eq:QMN})), the RR contribution can be accounted for  by the 
introduction 
of similar objects, $R_m$ and $R_{mn}$, with a group 
decomposition matching that of $Q$'s. The condition for supersymmetric vacua
will then reduce to algebraic equations of the form $R=-Q$.
This formulation is also more suitable to check the mirror symmetry 
proposal.

To write the analogue of (\ref{eq:QMN}) for non-zero RR fluxes we need 
to discuss in more detail the supersymmetry transformations  for type II 
theories 
and the general features of the solutions we are looking for.  

Our starting point is the democratic
formalism of  \cite{demo}, for which the 
supersymmetry  transformations for the gravitino and the dilatino are \footnote{Our convention
for $\sla F^{(2n)}$, differs from that in \cite{demo} in that we include a factor of $1/(2n)!$
in the definition of the slash 
(cf. Eq.(\ref{eq:clmap})).} 
\beq 
\label{eq:susyg} 
\delta \psi_M = D_M \epsilon + \frac{1}{4} H_M {\cal P} \epsilon + \frac{1}{16} e^{\phi}  
\sum_n  \sla \! {F_{2n}} \, \Gamma_{M} {\cal P}_n \, \epsilon  \, ,
\eeq 
\beq 
\label{eq:susyd} 
\delta \lambda = \left(\sla{\partial} \phi + \frac{1}{2} \sla \! H {\cal P}\right) \epsilon 
+ \frac{1}{8} e^{\phi}  
\sum_n (-1)^{2n} (5-2n)  \sla \! {F_{2n}} \,  
{\cal P}_n  \epsilon  \, .
\eeq 
$F_{2n}=dC_{2n-1} - H \wedge C_{2n-3}$ are the modified RR field strengths with
non standard Bianchi identities, that we will
call from now on simply RR field strengths, $n=0, \ldots,5$ for IIA and  
$n=1/2, \ldots,9/2$ for IIB and $H_M \equiv \frac{1}{2}H_{MNP} \Gamma^{NP}$.

 Note that the ``total" RR field 
involves both the  field strengths and their duals, and a self-duality 
relation is still to be imposed
\bea
F_{2n}  
= (- 1)^{Int[n]} \star_{10} F_{10-2n} \, .
\label{sd10}
\eea
The definitions of ${\cal P}$ and ${\cal P}_n$ are different in IIA and  
IIB: for IIA ${\cal P} = \Gamma_{11}$ and 
${\cal P}_n = \Gamma_{11}^n \sigma^1$,
while for 
IIB  ${\cal P} = -\sigma^3$, ${\cal P}_n = \sigma^1$ for $n+1/2$ even and 
${\cal P}_n = i \sigma^2$ for $n+1/2$ odd.
The two Majorana-Weyl supersymmetry parameters of type II supergravity are
arranged in the doublet  $\epsilon= (\epsilon_1,\epsilon_2)$.

It is useful to note that in the combination $\Gamma_M \delta \psi_M -
\delta \lambda$ 
the term corresponding to RR fluxes cancels. So instead of using the
gravitino  and the  dilatino equations, we will work  
with the gravitino and the modified dilatino equation
\beq
\Gamma^M \delta \psi_M - \delta \lambda = \Big(\sla \! {D} 
- \sla {\partial} \phi  + \frac14 \sla \! {H} {\cal P} \Big) \, 
\epsilon \, .
\label{eq:moddil}
\eeq

We are interested in solutions corresponding to warped compactifications to 4d
Minkowski space-time. So the 10d metric can be written as
\beq
\label{eq:metr}
ds^2_{10} = e^{2A(y)} d x_{\mu} d x^{\mu} + ds^2_6(y) \, ,
\eeq
and  we decompose gamma matrices, spinors and forms into 4d and 6d parts. 

We choose a Majorana representation for the 10d gamma matrices and we
split them according to
\beq
\Gamma_{\mu}= \gamma_{\mu} \otimes 1 \, , \ \ \Gamma_m=\gamma_5
\otimes \gamma_m \, ,
\eeq
where the 6d gammas are antihermitian and purely imaginary, and
\beq
\gamma_5= i \frac{1}{4!} \epsilon_{\mu \nu \lambda \rho} \gamma^{\mu \nu \lambda \rho} \, , \ \ \ 
  \gamma_7= -i \frac{1}{6!} \epsilon_{mnpqrs} \gamma^{mnpqrs} \, .
\eeq 

With the above choice for the 10d gamma matrices, we can now 
consider the decomposition of the spinors $\epsilon_i$.
In Type IIA the two supersymmetry parameters have opposite chirality, i.e. 
\begin{equation} 
\gamma_{11} \epsilon_1 = \epsilon_1 \, ,  \qquad   
\gamma_{11} \epsilon_2 = - \epsilon_2   \, ,
\end{equation} 
and  can be decomposed as  
\begin{eqnarray} 
\epsilon_1 & = &  \zeta_{+} \otimes \eta^1_{+} + \zeta_{-} \otimes 
\eta^1_{-} \, ,
\nonumber\\ 
\epsilon_2 & = & \zeta_{+} \otimes \eta^2_{-}  
+ \zeta_{-} \otimes \eta^2_{+} \, , 
\label{IIAans} 
\end{eqnarray} 
where $\zeta$ and $\eta^i$ are chiral spinors in 4 and 6 dimensions,
respectively. The Majorana condition implies also
$(\zeta_{+})^* =  \zeta_{-}$, $(\eta^i_{+})^{*} = \eta^i_{-}$. 

For IIB, the two $\epsilon$  have the same chirality and we choose it to be 
positive, which leads to the decomposition
\beq 
\epsilon_i = \zeta_{+} \otimes \eta^i_{+} + \zeta_{-} \otimes \eta^i_{-} \, , 
\eeq 
where again  $\zeta_{+}^{*} =  \zeta_{-}$, $\eta_{+}^{i*} = \eta_{-}^i$. 

Finally, we need to decompose the RR field strengths. In order to preserve 
4d Poincare invariance, they should be of the form 
\beq 
\label{eq:rrfs} 
F_{2n} = {\hat F}_{2n} + Vol_4 \wedge {\tilde F}_{2n-4} \, .
\eeq 
Here ${\hat F}_{2n}$ stands  for purely internal fluxes. 
The self-duality of $F_{2n}$  now becomes ${\tilde F}_{2n-4} = 
(-1)^{Int[n]} \star_6  {\hat F}_{10-2n}$, and allows to write the  
RR part of (\ref{eq:susyg})  in terms of the internal fluxes only.
From now on we will work with only internal fluxes, and drop the hats in $F$.

With the above decompositions, the supersymmetry conditions (\ref{eq:susyg}),
(\ref{eq:moddil}) reduce to a set of equations on the two spinors
$\eta^i_+$. Having two internal spinors would give a SU(2) structure. 
In
this paper we are interested in manifolds with SU(3) structure, which
is defined by a single spinor $\eta_+$.
Then we should find a way to relate the spinors 
$\eta^1_+$ and $\eta^2_+$ to the spinor $\eta_+$.  
If $\eta_+ $ is normalized  ($\eta_+^\dagger \eta_+ = \frac{1}{2}$), 
the most general way the $\eta^i_+$ and $\eta_+$ can be related 
is
\beq  
\label{eq:basissp}
\eta^1_+=a \eta_+  \, , \ \ \eta^2_+= b \eta_+  \, ,
\eeq 
where $a$ and $b$ are complex functions of the internal space.
Similarly, complex conjugate relations hold for the negative chirality spinors.
In order to be able to define the RR analogues of the $Q$'s, 
we must then express all the spinors
on the internal manifold in terms of the basis
\beq \label{eq:basis}
\eta_{\pm}, \, \,  \gamma^m \eta_{\pm} \, .
\eeq

Coming back to the supersymmetry conditions for IIA, they become 
\beq 
\alpha \sla {\partial}A \eta_+  
+ \frac{i}{4} e^{\phi} \sla \!{F}_{A1}\eta_- = 0 \, ,
\eeq
\beq 
\alpha D_m \eta_+ + ( \partial_m \alpha + \frac{1}{4} \beta \sla \! {H}_m ) \eta_+ 
+ \frac{i}{8} e^{\phi} \sla \!{F}_{A1} \gamma_m  \eta_- = 0 \, ,
\eeq
\beq 
\alpha \sla \! D \eta_+ + \Big[\alpha {\partial} (2A - \phi + log
\alpha) +\frac{1}{4} \beta \sla \! H \Big] \eta_+ =0 \, , 
\eeq
while for IIB they are 
\beq 
\left[\alpha \sla {\partial} A + \frac{i}{4} e^{\phi} \sla
  \!{F}_{B1} \right] \eta_+ = 0  \, ,
\eeq 
\beq
\alpha D_m \eta_+ + \left[ \partial_m \alpha -\frac{1}{4} \beta \sla \! 
{H}_m - \frac{i}{8} e^{\phi} 
\sla \!{F}_{B1} \gamma_m \right] \eta_+ = 0  \, ,
\eeq 
\beq
\alpha \sla \! D \eta_+ + \Big[\alpha \sla {\partial} (2A - \phi + log
\alpha) - \frac{1}{4} \beta \sla \! H \Big] \eta_+ =0 \, .
\eeq 
where we have introduced $\alpha = a+ ib$ and $\beta = a - ib$.

In both cases,
the first and second equations come from the spacetime and the internal 
gravitino respectively, while the third one
comes from the modified dilatino. 
The RR field strengths have been grouped in the following way:
\bea \label{defGA} 
- F_{A1}
&\equiv& \beta^* F_{0}+\alpha^* F_{2}+\beta^* F_{4}+\alpha^* F_{6} \, ,
\nn \\ 
F_{A2}
&\equiv& \alpha^* F_{0}+\beta^* F_{2}+\alpha^* F_{4}+\beta^* F_{6} \, ,
\eea 
\bea \label{defGB}
F_{B1}&\equiv& \alpha F_{1}-\beta F_{3}+\alpha F_{5} \, , \nn \\ 
- F_{B2}&\equiv &\beta F_{1}-\alpha F_{3}+ \beta F_{5}  \, .
\eea 
A second set of equations with $F_{A1} \rightarrow F_{A2}$, 
$F_{B1} \rightarrow F_{B2}$ 
and $\alpha \leftrightarrow \beta$ comes from the 
supersymmetry transformations of the second 
gravitino and dilatino. 

As for the NS case, the idea now is to reduce all the terms appearing in the 
supersymmetry variations to the spinor basis (\ref{eq:basis}), by 
decomposing
the action of a generic  gamma matrix $\gamma^{m_1, ... , m_n}$ 
on $\eta_{\pm}$.
This allows to treat the supersymmetry constraints for IIA and IIB 
on a common ground. The generic form of the equations is now
\bea
\delta \Psi_{m}: &&i({ Q}_m +R_m) \eta_+ + i(Q_{mn}+R_{mn}) \gamma^{n}
\eta_-  = 0 \, ,\nn\\
\delta \Psi_{\mu}:&&S \eta_- + (S_m+A_m) \gamma^m \eta_+ = 0 \, ,\nn\\
\delta \lambda: &&T \eta_- +  T_{m} \gamma^{m} \eta_+ =0 \, ,
\label{eq:susyRQ}
\eea
where $Q$'s, $T$'s and $A$'s contain the contribution from the geometry 
and the NS fluxes, and the $S$'s 
and $R$'s come from the RR fluxes.
This general structure holds for both IIA and IIB, even though the explicit 
form of $Q$, $S$ and $R$ is theory dependent. The actual computation of these 
coefficients can be found in  the Appendix. Here we will simply list the 
results.  For IIA we have
\bea \label{eq:RSIIA}
A_{m} &=& \alpha \partial_{m} A \, ,\\
S&=&\frac{i}{4} e^{\phi} (\sla \!F_{A1} \sei )_0 \, , \nn\\
S_{m}&=&\frac{1}{4} e^{\phi}  {\rm Re} \Big[
(\sla \!F_{A1}  \sla  \overline{\Omega})_{m} \Big] \, , \nn \\
Q_{m}&=& -i \de_m \alpha +\frac{1}{2} J_m^{\ \, n}\left(\alpha W_5-\alpha W_4 
\right)_n+\frac12 \beta H^{(3)}_m \, ,  \nn \\
Q_{mn}&=& {\rm Re} \Big[\frac{1}{2} 
(\alpha W_1 + 3 i \beta H^{(1)}) \bar{P}_{mn} -
\frac14\Omega_{mnp}(\alpha W_4 
+ i \beta H^{(3)})^p  \, ,  \nn\\ 
& & - \frac{i}{8} ( \alpha W_3 + i \beta H^{(6)})_{mn} 
+ \frac{i}{2} \bar{P}_m^{\, \, \, \,  p} \alpha W_{2 \, \, pn} \Big] \, ,  \nn\\
R_{m} &=&  -\frac{i}{8} e^{\phi}  
( \sla  \overline{\Omega} \sla \!F_{A1})_{m} \, ,  \nn \\
R_{mn}&=&\frac{1}{4} e^{\phi} {\rm Re}
\Big[- ( \sla \!F_{A1 \, m}  \sei)_n  
+ \frac{1}{2}(\sla \!F_{A1} \sei)_0 g_{mn} 
+ (\sla \!F_{A1} \sei)_{mn} \Big] \, , \nn \\
T&=& \frac{3}{2}\left( i \alpha W_1 - \beta H^{(1)}\right) \, ,  \nn \\
T_m &=&   \alpha \, \partial_m (2A - \phi + \log \alpha) +
\alpha\left[W_{4\,m} + \frac i2 J_m^{\ \, n}(W_5 - W_4)_n \right] -
\frac 12 J_{mn} \beta H^{(3)}_n \, , \nn 
\eea
where we have defined 
$\sla  F_m \equiv \frac{1}{k!} F_{m i_1 ... i_k} \gamma^{i_1 .. i_k}$.
Similarly for IIB we get 
\bea \label{eq:RSIIB}
A_{m}&=& \alpha \partial_{m} A \, ,  \\
S&=&\frac{1}{4} e^{\phi}  (\sla \!F_{B1} \sla \Omega)_0 \, ,  \nn \\
S_{m}&=&\frac{1}{4} e^{\phi}\, {\rm Re} 
\Big[(\sla \!F_{B1}  \seim)_{m} \Big] \, ,  \nn \\
Q_{m}&=& -i \de_m \alpha +\frac{1}{2} J_m^{\ \, n}\left(\alpha W_5-\alpha W_4 
\right)_n -\frac12 \beta H^{(3)}_m  \, , \nn \\
Q_{mn}&=& {\rm Re} \Big[\frac{1}{2} 
(\alpha W_1 - 3 i \beta H^{(1)}) \bar{P}_{mn} -
\frac14\Omega_{mnp}(\alpha W_4 
- i \beta H^{(3)})^p \, ,  \nn\\ 
& & - \frac{i}{8} ( \alpha W_3 - i \beta H^{(6)})_{mn} 
+ \frac{i}{2} \bar{P}_m^{\, \, \, \,  p} \alpha W_{2 \, \, pn} \Big] \, ,   
\nn\\
R_{m}&=& - \frac{1}{8} e^{\phi}  
( \seim \sla \!F_{B1})_{m} \, , \nn \\
R_{mn}&=& \frac{1}{4} e^{\phi} {\rm Re} 
\Big[ i ( \sla \!F_{B1 \, m} \sla \Omega)_n  - i
 (\sla \!F_{B1} \sla \Omega)_{mn} 
- i  \frac{1}{2} (\sla \!F_{B1} \sla \Omega)_{0}\, g_{mn} \Big] \, , \nn \\
T&=& \frac{3}{2} \left( i \alpha W_1  +  \beta H^{(1)}\right) \, ,  \nn \\
T_m &=&   \alpha \, \partial_m (2A - \phi + \log \alpha) +
\alpha\left[W_{4\,m} + \frac i2 J_m^{\ \, n}(W_5 - W_4)_n \right] +
\frac 12 J_{mn} \beta H^{(3)}_n  \, . \nn
\eea

For both theories the supersymmetry constraints on the second dilatino 
and gravitino yield
the same expressions with $\alpha \leftrightarrow \beta$, 
$F_{A1} \rightarrow F_{A2}$ and   
$F_{B1} \rightarrow F_{B2}$ for IIA and IIB respectively. 

Notice that $S$ and $T$ are just the flux and geometric parts of the 
superpotential.

\section{Conditions on supersymmetric vacua}
\label{sec:susy}

While the analysis of the implications of the supersymmetry
differential equations for pure spinors
is done better using a different approach and will be developed in the last section, 
writing the supersymmetry equations as in
(\ref{eq:susyRQ})
is better suited for analysis of specific backgrounds as it 
allows to check how 
the NS matrices $Q$ balance against RR matrices $R$ representation by 
representation.

In IIA, the RR sector consists of a zero- and a six-form with one
component each, and a two- and a 
four-form with 15 components each, making a total of 32 components. 
Under $SU(3) \in SO(6)$, the zero- and six-form are singlet, while the
two- and four- form decompose as $15  
\rightarrow 1 \oplus 3 \oplus {\bar 3} \oplus 8$.
In IIB, we of course  
have again a total of 32 components, however now they are distributed  
between one- and five-form (in the $3 \oplus {\bar 3}$ each) and a three-form  
($1 \oplus 3 \oplus 6$ + conjugates). 
They contribute to the supersymmetry equations through the tensors R's
and S's. 
As for the NS sector it is convenient  
to switch to a holomorphic basis and analyze the matrices $R_i$, $R_{ij}$  
and $R_{\bi j}$, so we need to look only at half of the components  
on each side. 
Differently from $Q$'s  
(see   (\ref{eq:qmnhol}) and (\ref{eq:QMN})) which have the same components in
IIA and IIB, some of $R$'s are not generic - there is no 6 appearing in IIA side,  
and no 8 on IIB side. 
However, $R$'s and $S$'s together have a total of 16  
components in both IIA and IIB. We can collect all  the representations in a table: 
 
\beq
\begin{tabular}{|c|c|c||c|c|c|} \hline 
\multicolumn{3}{|c||}{\rm IIA} &\multicolumn{3}{|c|}{\rm IIB} \\ \hline\hline 
$Q_i: 3$ &$R_i : 3$ &  $S_{i}: 3$ & $Q_i : 3$ & $R_i : 3$ & $ S_{i}: 3$\\ 
$Q_{ij}: 6 \oplus {\bar 3} $ & - & - & $Q_{ij}: 6 \oplus {\bar 3} $ &
$R_{ij}: 6 \oplus {\bar 3}$ & - \\ 
$Q_{\bi j}: 1 \oplus 8$  &$R_{\bi j}: 1 \oplus  8$ & $S:1$ & $Q_{\bi
  j}: 1 \oplus  8$   &  - &$S:1$ \\  
\hline \end{tabular} \label{list}  
\eeq

The first columns represent the NS sectors and are the same for IIA
and IIB~;  
the mirror  symmetry (\ref{eq:prop})  
exchanges the second  
and the third lines. The other two columns correspond to the RR
sector, and the mirror transformation is
\bea 
R_{\bi j}({\rm IIA})  
&\leftrightarrow& - R_{ij}({\rm IIB})\ \  (1 \oplus 8 \leftrightarrow   6 \oplus {\bar 3} ) \nn\\
R_i &\leftrightarrow& -R_{\bi} \nn\\
S_i &\leftrightarrow& S_i \nn \\ 
S &\leftrightarrow& S \ .\label{Rexchange}
\eea

Since  some representations are missing in $R$'s,  $Q_{ij}({\rm IIA})$ and $Q_{\bi j}({\rm IIB})$ must be  
zero by themselves.
This is the first hint of the fact that the integrability conditions on 
one of the pure spinors do not receive RR contributions, as we will show in detail in section \ref{sec:pure}.

Using Eqs.(\ref{eq:RSIIA})-(\ref{eq:RSIIB}) we will now give an
analysis of solutions to the supersymmetry conditions of IIA and IIB
on a manifold of $SU(3)$ structure. (These are necessary conditions;
Bianchi identities still have to be
imposed to make such backgrounds solutions.) As we will see,
one of the main difficulties in this analysis is that according to the relation between the 
normalizations $\alpha$ and $\beta$ of the two spinors, the equations we get can
be dependent or independent. 
It proves useful to reduce the
freedom of these two quantities by the following argument. For an
SU(3) structure, the two--form and three--form should satisfy 
$J \wedge \Omega=0$
and $i \Omega \wedge \bar \Omega =\frac{2}{3!}J^3$. Both relations are left
invariant if one redefines $\Omega\to \Omega e^{i\psi}$, with $\psi$
an arbitrary real function. This shifts $W_5\to W_5 +i d\psi$. In the
analysis below, we would always find such a spurious contribution to
$W_5$ to all solutions. This freedom can also be expressed as a
rescaling of the spinor $\eta_+\to e^{i\psi} \eta_+$, or as $\alpha\to
\alpha e^{i\psi}, \beta\to \beta e^{i\psi}$. In what follows, we fix
it by setting $ \mathrm{Arg}(\alpha)  +  \mathrm{Arg}(\beta)=0.$

\vspace{.3cm}


\vspace{.3cm}
\noindent
We start from type IIA theory and we derive the conditions
representation by representation.
\vspace{.3cm}

\underline{Scalars}: 
relations among them come from setting $S=0$, $Q_{\bi j}^{(1)}+
R_{\bi j}^{(1)}=0$ and $T=0$.
The last condition, $T=0$, imposes (remember that there is a second
set of tensors $T$, $R$, $S$, etc, obtained from
(\ref{eq:RSIIA}) by exchanging $\alpha$ with $\beta$, so $T=0$ gives
two equations): $i \alpha W_1 - \beta H^{(1)} =0$ and
$i \beta W_1 - \alpha H^{(1)} =0$. 
From these we see that the only Ansaetze that allow for nonzero
scalars in the torsion and H-flux are
$\alpha=\pm \beta$. We consider first the case $\alpha \neq \pm \beta$,
and we will analyze the case of the
equality later. When   $\alpha \neq \pm \beta$ we get $W_1=H^{(1)}=0$, 
which means $Q_{\bi j}^{(1)}=0$. 
The other two conditions,
 $S=0$ and $R_{\bi j}^{(1)}=0$, give four (complex) homogeneous equations 
%
%
for the remaining four (real) RR  scalars,
$F_{0}^{(1)}$, $F_{2}^{(1)}$, $F_{4}^{(1)}$ and $F_{6}^{(1)}$. 
These are four independent equations except when $\alpha=\pm \beta$. 
So for $\alpha \neq \pm \beta$, all scalars are zero.

The case $\alpha = \pm \beta$ works differently. As we showed, the
condition $T=0$ allows for nonzero $W_1$ and $H^{(1)}$
satisfying  $W_1 \mp i H^{(1)} =0 $. On the other hand, adding and
subtracting the equations coming from 
$Q_{\bi j}^{(1)}+ R_{\bi j}^{(1)}=0$ we can get rid of the RR piece,
and we get $ W_1 \pm 3i H^{(1)} =0$.
So for this particular Ansatz we also obtain $W_1=H^{(1)}=0$. 
But this Ansatz does allow for RR scalars if they are all equal
among them: $F_0^{(1)} = \pm F_2^{(1)} =F_4^{(1)} = \pm F_6^{(1)} $.

\underline{$8\oplus 8$}: 
Conditions for these
representations come from  $Q_{\bi j}^{(8)}+R_{\bi  j}^{(8)}=0$. 
$Q_{\bi j}^{(8)}$ is proportional to $W_2$, while $R_{\bi j}^{(8)}$
contains the non-primitive (1,1) piece of
the RR 2-form and the non-primitive (2,2) piece of the
4-form. 
These are four real homogeneous equations for four real variables 
($W_2^+$, $W_2^-$ , $ F_{2}^{(8)}$ and
 $ F_{4}^{(8)}$). The determinant of this system is proportional to 
$\mathrm{Re}(\alpha\bar\beta)\mathrm{Re}(\alpha^2 +\beta^2)$. Given 
the fixing of the total phase of $\alpha$ and $\beta$ that we did
above, the determinant is zero only for $\alpha/\beta$ purely
imaginary.
In this case, there is a solution $W_2^+= e^{\phi}
\frac{\mathrm{Im}(\alpha^2)}{|\alpha|^2} F_2^{(8)}$, $W_2^-=e^{\phi}
\frac{\mathrm{Re}(\alpha^2)}{|\alpha|^2}F_2^{(8)}$, which is a
variation on the holomorphic monopole in \cite{monopole}. For
$\alpha=i\beta$ another independent solution appears,
$W_2^+=F_4^{(8)}$. (Of course the two can be combined.) 
When we are not in any of these special cases, all the $8$ vanish; in
particular, $W_2$ is zero, which together with the condition $W_1=0$ obtained in the analysis for the
scalars, implies the manifold is complex. 

\underline{$6\oplus {\bar 6}$}: As 
it can be seen from table (\ref{list}),
IIA solutions should satisfy $Q_{ij}=0$, which means in particular
$Q_{ij}^{(6)}=0$. This gives again two homogeneous equations 
($(\alpha W_3+ i \beta H^{(6)})_{ij}=0$ and the same with
$\alpha$ and $\beta$
exchanged) that have nontrivial solution only when $\alpha=\pm
\beta$. So, for $\alpha \neq \pm \beta$,
$W_3=H^{(6)}=0$, 
while for $\alpha = \pm \beta$ we get $W_3=\pm *_6 H^{(6)}$.

\underline{$3\oplus {\bar 3}$}: 
$Q_{ij}^{(3)}=0$ sets $\alpha W_4+ i \beta H^{(3)}=0$, and the same
with $\alpha$ and $\beta$ exchanged. So again, for $\alpha
\neq \pm \beta$ both $W_4$ and $H^{(3)}$ are zero, while
for $\alpha = \pm \beta$ we get $W_4= \pm i H^{(3)}$.
For the latter, we get that all the RR vectors are zero
and $W_5=2W_4=\pm 2iH^{(\bar 3)}= 2 \bar\del\phi$, 
a condition familiar from \cite{hull,strominger}. 

The case in which $\alpha\neq \pm\beta$ is more intricate. Some of the
many equations can be recombined right away. In particular, one gets
that the ratio $\alpha/\beta$ is a {\it constant}. This fact is
strikingly different from the IIB case we will analyze next; we will
comment on this difference later. The remaining equations form a
system whose determinant is proportional this time to
$\mathrm{Re}(\alpha\bar\beta)$. As above, this vanishes for
$\alpha/\beta$ purely imaginary. The solution is in this case 
$F_2^{(\bar 3)}=\frac23 i\bar\partial \phi$, 
$F_4^{(\bar 3)}=0$, $W_4=0=H^{(3)}$, $W_5=\frac13 \bar\partial \phi$, 
$\bar\partial A =-\frac13 \bar\partial \phi$. These conditions are
those in \cite{monopole} again. 

\vspace{.3cm}

The table below is a summary of what we obtained. 
In the vector representation (the last row) we have only written the fluxes
that are not zero.

From the analysis of
the vectors we have that $\alpha/\beta$ is a constant. Depending on
what this constant is, we have different solutions. 
\begin{center}
\renewcommand{\arraystretch}{1.5}
\begin{tabular}{|c||c|c|c||}\hline 
{\bf IIA}& $\alpha=\pm\beta$ & \multicolumn{2}{|c||}{$\alpha=i
  k\beta$}\\\hline\hline
\multirow{2}{*}{1}&\multicolumn{3}{|c||}{$W_1=H^{(1)}=0$}\\ \cline{2-4}
&\minicent3{\vspace{.2cm}$F_0^{(1)} = \pm F_2^{(1)} =F_4^{(1)} = \pm F_6^{(1)} $\vspace{.2cm}} &
\multicolumn{2}{|c||}{$F_{2n}^{(1)}=0$}\\\hline
\multirow{2}{*}{8}&\multirow{2}{*}  &generic $k$ & $k=1$\\\cline{3-4}
&$W_2= F_2^{(8)}= F_4^{(8)}=0$&\minicent4{\vspace{.2cm}$W_2^+=e^{\phi}
\frac{\mathrm{Im}(\alpha^2)}{|\alpha|^2} F_2^{(8)}$ \\ $W_2^-=e^{\phi}
\frac{\mathrm{Re}(\alpha^2)}{|\alpha|^2}F_2^{(8)}$\vspace{.2cm}}& 
\minicent5{\vspace{.2cm}$W_2^+=e^{\phi}
\frac{\mathrm{Im}(\alpha^2)}{|\alpha|^2} F_2^{(8)}+e^{\phi}F_4^{(8)}$ \\ $W_2^-=
e^{\phi}\frac{\mathrm{Re}(\alpha^2)}{|\alpha|^2}F_2^{(8)} \ \ \ \ \ \ \ \ \ $}
\\\hline
6&$W_3=\pm *_6 H^{(6)}$ & \multicolumn{2}{|c||}{$W_3= H^{(6)}=0$ }\\\hline
3&\minicent3{\vspace{.2cm} $\bar W_5=2W_4=\pm 2iH^{(\bar 3)}=\bar\del\phi$\\
$\bar\del A=\bar\del \alpha=0$\vspace{.2cm}}&\multicolumn{2}{|c||}{
\minicent6{\vspace{.2cm}$F_2^{(\bar 3)}=2i \bar W_5=-2i \bar\partial A =
\frac23 i\bar\partial \phi$, $W_4=0$\vspace{.2cm}}}\\\hline\hline 
\end{tabular}
\end{center}

We see that essentially the only two supersymmetric solutions of
IIA are given by the common sector \cite{hull,strominger} and by the
holomorphic monopole \cite{monopole}, with some variant. There is
a natural way of seeing that these two should be the only allowed
classes. Let us start from supersymmetric M--theory
compactifications. It was shown \cite{kmt} that these are given by
seven--dimensional manifolds with SU(3) structure. This is
similar to SU(3) structure in six dimensions, but it also includes a
vector $v$. Reducing to six dimensions will involve
another vector $e^7$, which has to be Killing. In general, $e^7$ and
$v$ together give rise to an SU(2) structure in six dimensions, which
is not what we are considering here. If we want an SU(3) structure,
there are two possibilities. Either $v$ and $e^7$ actually coincide,
or the SU(3) structure in seven dimensions degenerates to a G$_2$
structure. The former case gives by reduction \cite{dap} the first
column of the table. In the latter case, a G$_2$ structure does not
actually allow for fluxes \cite{kmt}, that is, we are forced to G$_2$
{\it holonomy} manifolds. Upon reduction, this leads to the
holomorphic monopole geometry \cite{monopole} of the second column.

\vspace{.3cm}
\noindent
The same analysis can be repeated for the type IIB theory. 
\vspace{.3cm}

\underline{Scalars}: the condition $S=0$ sets $F_{3}^{(1)}=0$, while
$T=0$ together with $Q_{\bi j}=0$, 
which annihilate two different combinations of $W_1$ and  $H^{(1)}$,
set both of them to zero. 
The fact that all scalars are zero in supersymmetric IIB solutions has
already been noticed in \cite{Dallagata,Frey}. 

\underline{$8 \oplus 8$}: $Q_{\bi j}^{(8)}=0$ sets $W_2=0$. 
This condition can be easily
obtained just by noticing that there  is no $8$ representation in $H$,
and neither there is in the RR fluxes in IIB.    
$W_2=0$, together with the condition $W_1=0$, mean that in IIB the 
complex structure is always integrable. 
This conclusion, which was previously 
obtained in \cite{Dallagata,Frey}, is straightforward to get from (\ref{eq:RSIIB}). 

\underline{$6 \oplus {\bar 6}$}: 
the condition $ Q_{ij}^{(6)} + R_{ij}^{(6)}=0$ gives two complex
equations for three complex variables
$W_3$, $F_{3}^{(6)}$ and  $H_3^{(6)}$, so we can write two of them
as as a function of the third one:
\begin{eqnarray}
&&
(\alpha^2 - \beta^2)\, W_3 \,= e^{\phi} \,2 \alpha \beta \,F^{(6)}_{3} \nn \\
&& (\alpha^2 + \beta^2)\, W_3 \,= -2 \alpha \beta \,*_6H^{(6)}
\label{eq:6int}
 \\
&& (\alpha^2 - \beta^2)\, H^{(6)}= e^{\phi} \,(\alpha^2 + \beta^2)
*_6F^{(6)}_{3}\nn
\end{eqnarray}
 Only for the cases $\alpha=0$ or
$\beta=0$; $\alpha=\pm \beta$ and $\alpha = \pm i \beta$ 
one of the three vanishes ($W_3$ for the first two cases, while
$F_{3}^{(6)}$ and  $H_{3}^{(6)}$ for the last two respectively). These
three cases correspond to well known solutions (called respectively type
B, A and C in \cite{interpolating}), as we will discuss below.

\underline{$3 \oplus {\bar 3}$}: 
As opposed to IIA, we cannot isolate two equations which naturally
separate the case $\alpha=\pm\beta$ from the others. Neither we find
an equation imposing $\alpha/\beta= const. $. One thing that one can
do is to look at the three special cases singled out by looking at the
6, and thus imposing by hand $\alpha=\pm\beta$ (case A), $\alpha=0$ or
$\beta=0$ (case B), $\alpha=\pm i\beta$ (case C). In all these cases
we can analyze the most general solutions for the vectors. 

Another thing one can do is to look for solutions which correspond to
generic values of $\alpha$ and $\beta$ satisfying the ``gauge fixing'' condition 
$Arg(\alpha)+ Arg(\beta)=0$.
In this case all vector the components of the torsion and the fluxes, except
for $F_1$ are
non-zero and proportional to $\bar \del \beta$
\begin{equation}
  \label{eq:intvec}
\begin{array}{l}
e^{\phi} F^{(\bar 3)}_3 = 
\frac{-8 \alpha }{3\alpha^2 +  \beta^2}  \bar \del
\beta \, , \\
e^{\phi} F^{(\bar 3)}_5 = 
 \frac{ 4 i (\alpha^2 +\beta^2)}{\beta (3\alpha^2 +  \beta^2)}  \bar \del
\beta \, ,
\end{array} \quad
\begin{array}{l}
W_4 =  \frac{ 4 (\alpha^2 + \beta^2)^2}{\beta (\alpha^2 - \beta^2)
(3\alpha^2 + \beta^2)} \bar \del \beta \, ,  \\
\bar W_5 =  \frac{2( 3\alpha^2 + \beta^2)}{\beta(\alpha^2 - \beta^2)} \bar \del
\beta \, , \\
H^{(\bar 3)} = 
 \frac{-8i \alpha (\alpha^2 + \beta^2)}{(\alpha^2 - \beta^2)
(3\alpha^2 + \beta^2)} \bar \del \beta \, ,
\end{array}
\quad
\begin{array}{l}
\bar \partial  A = - \frac{2 (\alpha^2 - \beta^2)}{\beta(3\alpha^2 + \beta^2)}
\bar \del \beta \, ,  \\
\bar \del \phi =   \frac{16 \alpha^2 \beta }{ (3\alpha^2 + 
  \beta^2) (\alpha^2 - \beta^2)} \bar \del \beta  \, . 
\end{array}
\end{equation}

\noindent Moreover the functions $\alpha$ and $\beta$ 
are related to the warp
factor by $A = \log(|\alpha|^2 + |\beta|^2)$, as expected for a
supersymmetric IIB compactification \cite{Frey}.
This is an interpolating solution between type A and type B, similar to that in 
\cite{interpolating}. Indeed, although the explicit expression for the fields
are different due to the different choices for how to fix the total phase of
$\alpha$ and $\beta$, a straightforward computation shows that the
relations among the fields are actually the same.
Notice also that the ratios among the various fields, for example the one
between $W_4$ and $H^{(\bar 3)}$, or between $H^{(\bar 3)}$ and
$F_3^{(\bar 3)}$, are singular exactly for the special
values of $\alpha$ and $\beta$ selected above~: $\alpha = \pm \beta$
and $\alpha =0$ or $\beta =0$. 

\vspace{.3cm}
 
To conclude the analysis of IIB, we summarize the main features of the
three special cases A,B and C.
As in type IIA,
quantities not mentioned in the table in a given representations are
vanishing. The cases shown are this time
naturally singled out, but do not exhaust the possibilities.  
As we have argued above (and as it was found in 
\cite{interpolating}) there exist solutions that interpolate between the
ones shown.
 
\hspace{-1.2cm}
\renewcommand{\arraystretch}{1.5}
\begin{tabular}{|c||c|c|c||c||}\hline 
{\bf IIB}& $\alpha=\pm\beta$ (A) & $\alpha=0$ or $\beta=0$ (B)& 
$\alpha=\pm i\beta$ (C) & {\small interp.} \\\hline\hline
1&\multicolumn{4}{|c||}{$W_1=F^{(1)}_3=H^{(1)}=0$}\\\hline
8&\multicolumn{4}{|c||}{$W_2=0$}\\\hline
6&\minicent{2.5}{\vspace{.2cm}$F_3^{(6)}=0$\\ $W_3=\mp * H^{(6)}$\vspace{.2cm}}&
\minicent{3}{$W_3=0$\\ $e^{\phi}F_3^{(6)}=* H^{(6)}$}&
\minicent{3}{$H^{(6)}=0$\\ $W_3= \pm e^{\phi} * F_3^{(6)}$}&
(\ref{eq:6int})\\\hline
\multirow{2}{*}{3}& \multirow{2}{*}{
\minicent{2.8}{$\bar W_5=2W_4=\pm 2iH^{(\bar 3)}= 2 \bar\del\phi$\\
$\bar\del A=\bar\del \alpha=0$}
}&
\minicent{4.5}{\vspace{.2cm}$ e^\phi F_5^{(\bar 3)}= 
\frac23 i \bar W_5 = i W_4=-2i\bar\partial A =-4i\bar\partial \log \alpha$\\
$\bar \partial \phi =0$ \vspace{.2cm}} &
\multirow{2}{*}{\minicent{4.2}{$\pm e^\phi F_3^{(\bar 3)}=2i \bar W_5=$\\
$- 2i\bar\partial A =- 4i\bar\partial \log \alpha=$\\
$-i \bar \partial \phi$}}&
\multirow{2}{*}{(\ref{eq:intvec})}\\\cline{3-3}
&& 
\minicent{4}{\vspace{.2cm}$e^\phi F_1^{(\bar 3)}=2 e^\phi F_5^{(\bar 3)}=$\\ 
$i \bar W_5 = i W_4=i\bar\partial  \phi$\vspace{.2cm}}
&&\\\hline\hline
\end{tabular}

\vspace{.4cm}

Before moving on, let us identify the columns of this table. The first 
represents Strominger's solution \cite{strominger}. The second column
has two sub-cases. 
The first is a conformal rescaling
of a Calabi--Yau metric, with constant dilaton. Klebanov--Strassler solution 
for the deformed conifold falls into this class. 
\cite{KS,mariana,Gubser}. The second, if one chooses
$F_3^{(3)}=H^{(3)}=0$, corresponds to F--theory on a Calabi--Yau
four--fold. Case C is the
S--dual of the purely NS solution of case A, a well--known example of
which is Maldacena--Nu{\~n}ez solution \cite{mn}. The metric here is the
same as for the case A scaled by exp($\phi$), and $\phi \rightarrow -
\phi$, $H_3 \rightarrow - F_3$.

As already mentioned, differently from the table for IIA, here we have
presented just some special solutions rather than a classification.
There is another big difference between the two theories. In IIA, the
ratio $\alpha/\beta$ was constant, and
this is not so in IIB. (This ratio is indeed a non
constant function in interpolating solutions.) One might wonder how
can these differences be compatible with mirror symmetry. The answer
is that the freedom present in the spinor Ansatz in IIB has
to be reflected somewhere in IIA, but not necessarily in the spinor
Ansatz, given that we have not determined here how this maps under
mirror symmetry. The mirror of the A case (which is naturally singled
out in both theories) corresponds to the same Ansatz on the IIA side. But 
the mirror of type B class does not neccesarily have the same Ansatz in IIA, 
and neither does that of type C. 
Both B and C may be mapped after all to the same Ansatz (the same
would then be true for the interpolating solution). A rough argument
for this comes comparing with branes on Calabi--Yau manifolds. 
A D3
brane extended over Minkowski space is in class B, whereas a D5 falls
in class C. These two wrap a point and a two--cycle respectively. In a
Calabi--Yau, both these branes would be mapped by mirror symmetry to a
brane wrapping a Special Lagrangian three--cycle and therefore
the corresponding source is a
D6. These are described by the monopole background which is so clearly
singled out in second column of table IIA.

Following the transformations of the $Q$'s is the practical
criterion for determining the mirrors. We observe that other than the 
$\alpha=\pm \beta$ solution for IIB, all the others have $Q_{ij}\neq 0$. 
Thus, they require a mirror with non--integrable complex structure.  
The fact that on type IIA side there is a unique solution 
leaves us no choice but to conclude that the monopole geometry is
the universal mirror for all IIB solutions. Given that for the former  
$dJ=0$, and thus the geometry is symplectic, it is a rather natural mirror
to the IIB geometry, which is always complex.

 
\section{Pure spinor equations} 
\label{sec:pure}

In this section we will finally derive the pure spinor equations
promised in the introduction, and thereby justify the title of the
paper. First of all we will introduce the formalism of \clss\ spinors,
to justify the use of the expression $e^{i J}$ often mentioned above
(see for example (\ref{eq:str})).
Then we will give a brief
mathematical introduction to the use of pure spinors in the context
of generalized complex geometry. This will motivate us to look for
differential equations on the pure spinors defined on a SU(3)
structure manifold, which we finally present in equations (\ref{p1}--\ref{p4}).

\subsection{Pure spinors}

We start by introducing the formalism of pure spinors. 
We will also comment on how to obtain torsions from them.

\clss\ is an algebra of matrices $\lambda^m, \rho_n$ that obey
$$
\{ \lambda^m, \lambda^n\} =0\ , \qquad 
\{ \lambda^m, \rho_n\} = \delta^m_n \ , \qquad
\{ \rho_m, \rho_n \}=0\ .
$$
We have chosen two different symbols, $\lambda$ and $\rho$, instead of the 
more commonly used $\gamma^m$ and $\gamma_m$,  to emphasize that these matrices
are independent, they cannot be obtained from 
each other by raising and lowering
indices with the metric. So the number of gamma matrices is twice the 
dimension of
the manifold, in our case twelve. 
The representation of this algebra which is usually
taken, and to which we will stick, is on the vector space of formal sums of
forms of all degrees, $\oplus_{i=1}^6 \Lambda^i T^*$. Then $\lambda^m= dx^m
\wedge$, and $\rho_n = \iota_{\partial_n}$ (see footnote \ref{iota} for the explicit action of $\iota_{\partial_n}$). 

A pure spinor is one whose annihilator is a six--dimensional space in the
twelve--dimensional algebra \clss. On an SU(3) structure manifold there are
two natural pure spinors. One is simply $\Omega$, which is annihilated by 
$\lambda^i$ ($  \Ox_{jkl} \, dz^i \wedge dz^j \wedge dz^k \wedge dz^l =0$) and 
$\rho_{\bi}$ ($\Ox_{\bi kl}=0$). Another, which might seem more exotic, is 
$e^{i J} \equiv 1 + i J -1/2 J\wedge J -i/6 J\wedge J \wedge J$. It is
annihilated by $\rho_m + i J_{mn} \lambda^n$, as it is easy to check using
$J_m\,^n J_n\,^p = -\delta_m \,^p$. 

We will also use the familiar fact that we can map a form
(or a formal sum of them) to an element of the {\it usual} Clifford algebra, 
Clifford(6):
\begin{equation}
  \label{eq:clmap}
C\equiv\sum_k \frac{1}{k!}C^{(k)}_{i_1\ldots i_k} dx^{i_i}\wedge\ldots\wedge dx^{i_k}\qquad
\longleftrightarrow\qquad
\sla C \equiv
\sum_k \frac{1}{k!}C^{(k)}_{i_1\ldots i_k} \gamma^{i_i\ldots i_k}_{\alpha\beta} \ .
\end{equation}
An object in Clifford(6) can also be seen as a bispinor, since it has two free
spinor indices. So we have realized \clss\ spinors as bispinors, which are
more useful in string theory.
We will see that it is crucial that $e^{i J}\!\!\!\!\! 
\begin{picture}(10,10)
\put(0,0){\line(1,1){10}}
\end{picture}
$
and $\sla \Omega$ can be reexpressed in terms of tensor products of $\eta$, as
stated in the introduction. 
Another useful technical fact is that one can realize
$\lambda$ and $\rho$ also as combinations of the more familiar $\gamma$'s 
acting on the left and on the right of a bispinor. For example, 
$\lambda^m C^{(k)} \longleftrightarrow$ 
$\frac{1}{2}(\gamma^m {\sla C^{(k)}} \pm {\sla C^{(k)}} \gamma^m)$ 
when the plus (minus) sign corresponds to  $k$ even (odd).

All this technical machinery is mainly needed to give a meaning to the
expression $e^{iJ}$. We have just seen that it is a pure spinor, as it
is $\Omega$. Thus on a manifold of SU(3) structure there
are always two pure spinors. In this formulation, it is not unnatural
to think there might be a mirror symmetry exchanging these two:
$$
e^{iJ} \longleftrightarrow \Omega\ .
$$
This was indeed the formulation of mirror symmetry in \cite{fmt}. In
the first part of this paper, we have lent credence to this
conjecture by showing that supergravity (its supersymmetry
transformations) can be rewritten in terms of these two pure spinors
alone (or rather under their bispinor counterparts, eq. (\ref{eq:clmap})).
We have also given explicitly the exchange under which the two are
symmetric, eq. (\ref{Rexchange}). Later in this section we will also show
how
the supergravity equations imply differential equations in which they
appear symmetrically. For now, let us comment a moment about what is
the interpretation of the $R_{mn}$
from the point of view of pure spinors. 

In the Appendix  we use Fierz identities crucially, see
for example equation (\ref{eq:fz}). In that conventional treatment,
one uses a basis $\gamma^{m_i\ldots m_k}$ for expanding an arbitrary
element of the Clifford(6) algebra. These are obtained from a trivial
vacuum 1 acting with all the possible gamma matrices. However, this
procedure can be repeated just as well replacing 1 with another pure
spinor. In mathematical terms, Gualtieri \cite{gualtieri} introduces
(section 3.6) a
filtration of Clifford algebra by the number of gamma matrices acting
on a given pure spinor. In a more concrete language, this yields
another basis for the Clifford algebra. For example, 
with $\Omega$ as a Clifford vacuum, the basis is given by  
$\gamma_{i_1\ldots i_k} \Omega \gamma_{j_1 \ldots j_l}$: all the possible
holomorphic gammas from each side of $\Omega$. 
More explicitly, the first few read 
$\Omega$, $\gamma_i \Omega$, $\Omega \gamma_i$, $\gamma_i \Omega
\gamma_i$. Analogously, taking $e^{iJ}$ as a Clifford vacuum, 
the basis is $\gamma_{\bar i_1\ldots \bar i_k} e^{iJ} \gamma_{j_1 \ldots j_k}$. 
One can use these bases equally well to derive Fierz identities. For
the usual basis $\gamma^{m_1 \ldots m_k}$, the coefficients of the
expansion of a bispinor $F$ are $\mathrm{Tr}(F \gamma_{m_1\ldots m_k})$.
(If instead of $F$ we have $\eta_+\otimes
\eta_\pm^\dagger$, then we find (\ref{eq:fz})). If we now use one of
our new bases, say the one relative to $\Omega$, the first
coefficients of the expansion would now look
$$
\mathrm{Tr}(F\bar\Omega)\ , \quad \mathrm{Tr}(F\gamma^i \bar\Omega)\ , \quad
\mathrm{Tr}(F \bar\Omega \gamma^i)\ , \quad \mathrm{Tr}(F\gamma^i
\bar\Omega\gamma^j)\ .
$$
These, along with their analogues built using $e^{iJ}$, are nothing
but the $S$, $S_m$, $R_m$, $R_{mn}$ introduced in the previous
sections, as detailed in the Appendix. This gives us an
intuitive explanation of the exchange (\ref{Rexchange}). Since the pure
spinors are exchanged, $e^{iJ} \leftrightarrow \Omega$, all the tower
of states built by Clifford action from them will be exchanged (the
``filtration''); the coefficients of the expansions will be exchanged
too, and this is what (\ref{Rexchange}) provides.

Let us finally also comment on how to get torsions from pure
spinors. Schematically one has 
\begin{equation} 
  \label{eq:torscldd} 
  D_m e^{i J} = 
Im( q_m^{(2)}\cdot \Omega)\ ,\qquad 
  D_m \Omega = -i q_m\Omega - Im(q_m^{(2)}\cdot e^{i J}) \ . 
\end{equation} 
In these equations, $q_m^{(2)}\cdot$ is the usual 
Clifford product of $q_{mn}$ using only second index.

\subsection{Pure spinors in generalized complex geometry}
\label{sec:h}


We will now see how pure spinors are used in the context of
generalized complex geometry.  
The basic thing we want to show 
is that
pure spinors can be used instead of {\it generalized almost complex
structures}.

The idea behind generalized complex geometry is to 
consider the direct sum of the tangent and cotangent bundle, 
rather than the tangent bundle itself, and to generalize the 
standard machinery of complex geometry.

If ordinary almost complex structures $J$ are bundle maps from $T$
to itself that square to -1, {\it generalized} almost complex structures
${\cal J}$ are maps of $T \oplus  T^*$ to itself that square to -1 
\footnote{As for an almost complex structure, ${\cal J}$ must also satisfy the 
hermiticity condition ${\cal J}^{t} {\cal I} {\cal J} = {\cal I}$, 
with the respect to the natural metric on $T \oplus  T^*$, 
${\cal I}={{0 \ \ 1} \choose {1 \ \ 0}}$.}
\beq
{\cal J} = 
\left(
\begin{array}{cc}
J& P\\ L & K 
\end{array}
\right) \, ,
\eeq
where $J : T{\cal M} \rightarrow T{\cal M}$,  
$P : T^*{\cal M} \rightarrow T{\cal M}$,  
$L : T{\cal M} \rightarrow T^*{\cal M}$ and  
$K : T^*{\cal M} \rightarrow T^*{\cal M}$.

From this expression it is easy to see that usual complex structures are
naturally embedded into ${\cal J}$: they correspond to the choice 
\beq
{\cal J}_1 \equiv 
\left(
\begin{array}{cc}
J & 0 \\ 0 & -J^t 
\end{array}
\right) \, ,
\eeq
with $J_m^{\, n}$ an almost complex structure. 
Another example of generalized almost complex structure can be built
using a non degenerate 
two--form $\omega$, 
\beq
{\cal J}_2\equiv
\left(
\begin{array}{cc}
0 & -\omega^{-1} \\\omega & 0 
\end{array}
\right) \, .
\eeq

There is a
natural integrability condition for generalized almost complex
structures, analogous to the
integrability condition for usual almost complex structures.
For the usual complex structures integrability, namely the vanishing of the 
Nijenhuis tensor (or in our terms, $q_{i \bj}=0$), 
can be written as a condition on the Lie bracket on $T$.
For generalized almost complex structures the Lie bracket is replaced by a 
certain bracket on $T \oplus T^*$, called the {\it Courant} bracket, 
which does not satisfy Jacobi in
general, but which does on the $i$--eigenspaces of a ${\cal J}$.

In case
these new conditions are fulfilled, we can drop the ``almost'' and speak of
generalized complex structures. It is interesting to see what this condition
is for the two examples above. For the one we called ${\cal J}_1$,
which was built from an almost complex structure, integrability coincides 
with the ordinary meaning, thus making it a complex structure.
For ${\cal J}_2$, which was built from a two--form $\omega$, the condition 
becomes $d\omega=0$, thus making $\omega$ into a {\it symplectic} form. 
%

These two examples are not exhaustive, and the most general
generalized complex structure {\it interpolates} between 
complex and symplectic manifolds. It is immediately clear that this formalism
then must be useful for mirror symmetry: although for the physical string
mirror symmetry is an exchange of \cy s, for the topological string it can be
formulated as sending symplectic manifolds into complex ones, and vice versa.

Now, what is
more immediately relevant 
in our  context is that generalized
complex geometry can be reformulated in terms of pure spinors. First of all
there is an algebraic correspondence between a generalized almost complex
structure ${\cal J}$ and a pure spinor $\varphi$. As an example, 
${\cal J}_1$ above is sent by this
correspondence to a section of the bundle of $(3,0)$ forms, which in the case 
of SU(3) structures exists: we have called it $\Omega$ so far.
The other example, ${\cal J}_2$  is sent to $e^{i J}$, where we have
renamed the symplectic form $\omega$ as $J$, as in the rest of the paper.

Under this correspondence, the integrability condition for a generalized almost
complex structure is equivalent to the condition that the spinor $\varphi$ 
is closed. This means that every degree of $\varphi$ is separately closed
(remember that a \clss\ spinor is a formal sum of forms).\footnote{More
  precisely, the condition is that there exist a vector $v$ and a one--form
  $\xi$ such that $d\varphi= (v\llcorner + \xi ) \varphi$.\label{foot:vxi}} 
Manifolds on which a
closed pure spinor exists were called generalized \cy s by Hitchin
\cite{hitchin}. 

There is also a possibility of adding a three--form $H$ to the story. 
Using a three--form, the Courant bracket can be modified, and with it 
the integrability
condition. Not surprisingly, this also corresponds to a modification of the 
condition on the pure spinor, which now becomes 
\begin{equation}
  \label{eq:Hw}
(d + H\wedge) \varphi=(v\llcorner + \xi ) \varphi\   
\end{equation}
for some $v$ and $\xi$ (compare with footnote \ref{foot:vxi}).
If we decompose $\varphi$ in forms, $\sum \varphi_{(k)}$, the
condition means this time that $d \varphi_{(k)} + H\wedge \varphi_{(k-2)}=v\llcorner \varphi_{(k+2)} + \xi \wedge \varphi_{(k)}$.

\subsection{Supersymmetry equations for pure spinors}

In this section we will finally use the work done on supergravity to derive
equations on the two pure spinors. These equations do not encode all the
information coming from the supersymmetry conditions. They are rather the
counterpart of the internal gravitino, in that they encode derivatives of
objects that can be used to define the structure. They capture the information
about the intrinsic torsion of the manifold; but in general as we have seen
there are more conditions, equaling components of fluxes (and derivatives of
the dilaton and warping) among each other. 
A natural question for a string theorist is whether (\ref{eq:Hw}) is relevant
in any way to compactifications with fluxes. The first natural example to look
at is of course the case with only $H$  present \cite{hull,
  strominger}
In section \ref{sec:h} we have already
noticed that the manifold is complex; with little more effort,
one gets that $e^{2\phi} \Omega$ is closed. So in that case the manifold is
generalized complex. 

In this, however, $H$ did not enter. It is natural to guess that the
condition on $e^{iJ}$ will be involving $H$, and maybe even that it would be
as in (\ref{eq:Hw}). 

Unfortunately, this guess turns out to be incorrect. One way to see this is
to use the so--called torsional equation
which can be seen to follow from the equations on the intrinsic torsions (for
example). This reads $\del J =-i H^{(2,1)}$.  Then we have $d e^{i J} =
\left(H^{(2,1)} - H^{(1,2)}\right) e^{iJ}$, falling short of proving
(\ref{eq:Hw}) for $e^{iJ}$. 

We may, then, propose another point of view. Hitting (\ref{eq:str}) with an
extra $\lambda$ (a wedging), one gets an equation 
\begin{equation}
  \label{eq:modHw}
 (d+ \frac14 H_{mnp}dx^m dx^n \iota^p) \varphi =0\ .
\end{equation}
So, $H$ acts by contraction of one index and wedging of the other two. Notice
also 
that this operator raises the degree of the forms in $\varphi$ by one, the same as $d$. 
This form of the equations has then the advantage that $H$ acts in the same way as $d$, which is not
the case in (\ref{eq:Hw}). 
But it would seem at this point that we have many options, and it is not clear
how to pick the most relevant or interesting. 

Introduction of RR fluxes makes the story very different. A priori, we might
again consider many combinations and get many different ways for $H$ to act. 
However, we have found 
 only one choice for which the RR flux only contributes
to one of the two spinor equations.
Here is a schematical 
description of how this computation is performed, for the pure spinor 
$e^{-iJ}$. 

Our strategy is to use the Clifford map (\ref{eq:clmap}) 
between pure spinors and bispinors. As we mentioned in the introduction, the 
pure spinors have a particularly simple form under this map. For example, 
$e^{-iJ}=8\eta_+\otimes\eta_+^\dagger$, see eq.(\ref{Fierz}).
The exterior derivative  $d(e^{-iJ})$ can be re-expressed in
the bispinor picture as the anticommutator 
$$
\{ \gamma^m, D_m(\eta_+\otimes \eta_+^\dagger) \} \  .
$$
The covariant derivative here is meant to be a bispinor covariant derivative, 
which corresponds to the ordinary covariant derivative of forms under the 
Clifford map, and which anyway reduces to exterior derivative when we fully 
antisymmetrize, as usual. To compute this object, one can use Leibniz rule 
for the covariant derivative of the bispinor, reducing it to 
$\{ \gamma^m, D_m(\eta_+)\otimes \eta_+\}$ plus its complex conjugate. 
The covariant derivative of the spinor con now be read off (\ref{IIAintf}) in 
IIA or (\ref{IIBintf}) in IIB. 
Actually, when  $\gamma^m$ acts on the left, one 
reconstructs the Dirac operator, which is better to read directly from
(\ref{IIAintf}) or (\ref{IIBintf}) in IIA and IIB, respectively.  
Substituting gives
\begin{eqnarray*}
\mathrm{IIA:}&& - [\sla \partial (2A -\phi +\log \alpha)
  +\frac\beta{4\alpha}\sla H]
\eta_+\otimes\eta_+^\dagger - (\partial_m \alpha +\frac\beta{4\alpha}H_m)
\eta_+\otimes\eta_+^\dagger \gamma^m \, ,\\
\mathrm{IIB:}&& - [\sla \partial (2A -\phi +\log \alpha)
  -\frac\beta{4\alpha}\sla H ]
\eta_+\otimes\eta_+^\dagger 
- (\partial_m \alpha -\frac\beta{4\alpha}H_m + \\
&& \hspace{7cm} -\frac i{4\alpha}e^{\phi} \sla F_{B\, 1}\gamma_m)
\eta_+\otimes\eta_+^\dagger \gamma^m \, ,
\end{eqnarray*}   
plus again the complex conjugates. 
Notice that in IIA $F$ has disappeared. This is because it would have
multiplied $\gamma_m \eta_-\otimes \eta_+^\dagger \gamma^m$. This
expression is zero because $\eta_-\otimes \eta_+^\dagger=-\frac
i8\sla\bar\Omega$, (see again (\ref{Fierz})) 
and $\gamma_m \gamma^{npq} \gamma^m=0$ in
six dimensions. This technical circumstance is what allows us to
make $F$ disappear in one of the pure spinor equations for both IIA
and IIB. As for the other expression, 
$\gamma_m \eta_+\otimes \eta_+^\dagger \gamma^m$, it can be
re-expressed in terms of $\eta_-\otimes \eta_-^\dagger$, but it is not
vanishing. 

It is now only required to go from the bispinor
picture  back to the form picture,  inverting the Clifford 
map (\ref{eq:clmap}). 
Analogous computations can be performed for $d\Omega$. 
The final results are as follows. In type IIA we have
\bea
e^{-f} d \Big(e^f e^{i J} \Big) &=&  -\frac{1}{2}
\frac{Re(\alpha\bar\beta)}{|\alpha|^2 + |\beta|^2} H\bullet e^{iJ} \, ,
\label{p1}\\
e^{-g} d \Big(e^g \Omega\Big) &=& 
-\frac14\frac{\beta^2+\alpha^2}{2\alpha{}\beta{}} H\bullet \Omega + \label{p2} \\
&-& \frac{e^\phi{}}{16}\frac1{2\alpha\beta} \Big( F \cdot
(-\frac14 e^{-i J} +1 + i\mathrm{vol}) - (-\frac14 e^{i J} +1 -i \mathrm{vol}) \cdot F^* \Big) \,
, \nonumber
\eea  
and in type IIB 
\bea
e^{-f} d \Big(e^f e^{iJ}\Big) &=& 
\frac12 \frac{Re(\alpha\bar\beta)}{|\alpha|^2 + |\beta|^2} H\bullet
e^{iJ} +
\label{p3} \\
&-& i\frac{ e^\phi{}}{16} \frac1{|\alpha{}|^2 + |\beta{}|^2}\Big( F\cdot  
(-\frac14e^{-i J} + 1 +i \mathrm {vol}) - (-2e^{-i J} + 1 +i \mathrm{vol}) \cdot F 
\Big) \ , \nonumber \\
e^{-g} d \Big(e^g \Omega \Big) &=&  \frac14\frac{\beta^2+\alpha^2}{2\alpha\beta{}} 
H\bullet \Omega{} \label{p4}\, .
\eea 
In both cases $f=2A-\phi+\log(|\alpha|^2 + |\beta|^2)$ and $g=2A-\phi +
\log(\alpha\beta)$, and $F \equiv (|\alpha|^2 - |\beta|^2)F_+ +
(\alpha\bar\beta-\bar\alpha \beta)F_-$, where $F_+$ is the hermitian
piece of the RR total form ($F_+=F_0+F_4$ in IIA, $F_+=F_1 +F_5$ in IIB) and
$F_-$ is the antihermitian piece ($F_-=F_2 +F_6$ in IIA and $F_-=F_5$ in
IIB). Since we wrote these equations with  forms rather than
bispinors, we explicitly denoted the Clifford product between forms  by
a $\cdot$; vol is the volume form, whose
image under the Clifford map would be $i \gamma$. 
 The operator $H\bullet$ is the same for all equations
and is defined by
\begin{equation}
  \label{eq:Hop}
H\bullet \equiv  H_{mnp} \left(dx^m dx^n \iota^p -\frac13
  \iota^m \iota^n \iota^p \right)\ .
\end{equation}

Although the RR piece is not very nice, it has a similar form in both
cases too. Most importantly, the action of the NS sector is always the
same. If the symmetry between the two theories was more or less guaranteed by
the analysis of the previous sections, nothing a priori guaranteed that the
operator $H\bullet$ would be the same also for both pure spinors, if not a
vague analogy with equation (\ref{eq:modHw}). Notice that in that case we had
been driven to a different choice of operator. But it was not the only one we
could have put in an (\ref{eq:modHw}), as we see now from specializing
equations (\ref{p1}--\ref{p4}) to the case with no RR.

Given the mathematical
discussion, it is natural to wonder if the operator $H\bullet$ we found has a
realization in terms of a twisting of the Courant bracket. Note
however that the combination $d + H\bullet$ does not square to zero
this time. 

With this caveat (or technical clarification) in mind, we will 
call any action of $H$-flux a twisting. Then it is easy to tell what
the main outcome of the equations (\ref{p1}--\ref{p2})  for IIA and 
(\ref{p3}--\ref{p4})  for IIB is. In each case, we have
one pure spinor equation that contains an exterior derivative and
$H$-twist. Thus, having a twisted closed pure spinor, or in other words
twisted generalized Calabi-Yau, is a necessary condition for having an
${\cal N} = 1$ vacuum. All the backgrounds constructed so far satisfy this
condition.

Strictly speaking we have proven this statement only for manifolds with
$SU(3)$ structure since the existence of this structure was assumed in
writing the covariant derivatives on the supersymmetry parameter. We would
like to argue though that this simplifying technical assumption can be
dropped and the result may hold in more general cases.  Going beyond the
$SU(3)$ structure is not a self-purpose - for example both for IIA and for
IIB there are known vacua which correspond to compactifications on
non-spin (yet indeed generalized Calabi-Yau) manifolds and the structure
group for these will not fit into $SU(3)$. To see the validity of the
formulae for the $U(3)$ structure case, we can go back to
(\ref{eq:newmirror}) and the footnote 7~: reformulating 
our results in
terms of the covariant derivatives of the fundamental form $J$ is the best
way to see that at least conceptually one can do without $\Omega$ or a
well-defined $\eta$. Note that treating $W_5$ not as a well-defined
one-form but as a connection would still allow to write (\ref{eq:deps}) making
use of the spin$^c$ structure of the manifold. We leave a more thorough
discussion of global (and non-geometric) aspects for future work.

A comment is due about the equivalence of (\ref{eq:susyRQ}) with the pure 
spinor equations. Clearly most of the information is contained in the part 
related to the internal gravitino and as already discussed this is 
captured by  (\ref{p1}--\ref{p4}).  The two vector-like equations simply 
serve to define the warp factor and the dilaton in terms of the geometry 
and flux data and then have to be added to the two pure spinor equations. 
As already noticed the remaining conditions $S=0=T$ comes from the
geometric and flux contributions to the superpotential. We conclude this
section by noticing that collecting all  pieces, the superpotential may
indeed be written using pure spinors (for the derivation see \cite{lg}):
$$
W=S+T = e^{\phi} e^{-B} \varphi_1 d(e^B (e^{-\phi} \varphi_2 +
iC)).
$$
The RR gauge field $C$ here stands for the formal sum of all forms 
and its rank 
being odd or even is correlated with the rank of $ \varphi_2$. In other 
words, $ \varphi_1 = {\rm exp }(iJ)$ and $ \varphi_2 = \Omega$ for IIA and 
the other way around for IIB. It is not hard to verify explicitly that 
this expression contains all the known contributions to the superpotential 
and nothing else.

\subsection{Topological strings and auxiliary fields.}

It is now natural to wonder, what the physical meaning of the
decoupling of one of the two pure spinors from the RR fields is. 

A natural answer can be found in the context of topological strings. 
It is well--known that the A and B model see only the symplectic and
the complex structure of a Calabi--Yau, respectively. We have seen in 
section \ref{sec:h},
in the context of generalized complex geometry,  
that the pure spinor $e^{i J}$ corresponds to the symplectic
structure, and the pure spinor $\Omega$ corresponds to the complex
structure. So, in our language topological strings only see one of the two
spinors.

If we have a nonlinear sigma model with a manifold of SU(3) structure
$M$ as target, the requirement  of extended supersymmetry will impose
certain differential conditions on the structure. Analogously, trying 
to define A and B models on $M$ will lead to certain
differential conditions on the pure spinor $e^{iJ}$ or $\Omega$
respectively, via the requirement of BRST closure. (With non-zero $H$,
this has not been done explicitly so far. 
The proper way of doing this is
most probably  using again generalized complex geometry and a framework
similar to \cite{lmtz}.)

Let us now try to switch RR fields on. It was argued in \cite{vafa} 
that their introduction does not modify topological string amplitudes.
(The
argument is also interesting in the present context, and we will come
back to it shortly.) If the topological model does not feel the RR
fields, the differential conditions we are finding  should not
change.
This is indeed what we have in equations (\ref{p1}) and 
(\ref{p4}). 

Let us now come back to the argument that guarantees that topological
strings are not affected by RR fields. It goes roughly as
follows. 
Suppose for a moment that we are on a Calabi--Yau, but
with RR fields switched on. The superpotential in this case is the
usual Gukov-Vafa-Witten one \cite{GVW, Gukov}. It was pointed out in
\cite{vafa} that
fluxes can be introduced by simply giving vacuum expectation value to
the auxiliary fields of the vector superfield. From this, one can also
see, by integration over one linear combination of the two $\theta$ of
${\cal N}=2 $ superspace\footnote{This choice will correspond to the
  $\alpha$ and $\beta$ in our Ansatz above.}, that the Gukov-Vafa-Witten
superpotential is reproduced {\it without changing the prepotential}
${\cal F}_0$. Since the latter is a topological amplitude, this means
topological amplitudes are not changed by RR fields. This argument was
made more precise in \cite{lmg} for IIB theory.

What is of more interest here, is that intrinsic torsions can also be
realized (analogously to RR fluxes) by giving expectation values to
auxiliary fields. This allows to reproduce the extension of the 
Gukov-Vafa-Witten
superpotential including intrinsic torsions that appeared in
\cite{glmw}. More precisely, the auxiliary fields for the vectors in type
IIA were argued in \cite{vafa} to contain torsions in the
representations 8 and 1, namely $W_1$ and $W_2$. On the other side,
the auxiliary fields for the vectors in type IIB were shown in \cite{lmg}
to contain torsions in representations 6 and 3, namely $W_3$ and
$W_4$. The effective four--dimensional theories should be equal
for two mirror compactifications. Thus we are lead to say that mirror
symmetry should exchange torsions in 6+3 with torsions in 8+1. This
is indeed consistent with the recipe given in \cite{fmt} and
reviewed here. 

Moreover, in \cite{lmg} a three--form superfield is introduced, which
has as auxiliary fields $W_3$ and $W_4$ (as well as fluxes). Its
lowest component  is simply $\Omega$ and it is given by 
$$
\Omega(\theta, {\tilde \theta}) = \Omega + \theta^2 (dJ+ H) + {\tilde
\theta}^2 (dJ-H) + \theta {\tilde \theta} (F - C_0 H) + \ldots
$$

The logic of the present paper tells that a similar object should exist for IIA
except that this time it is an $e^{iJ}$ superfield - namely an object
whose lowest component is given by a formal sum of terms. Note that such a
structure is also consistent with 
the flux superpotential
for IIA. Moreover the general intuition from mirror symmetry would tell
that in $\theta {\tilde \theta}$-term it is more natural to expect  
an NS contribution from the metric rather than the $H$-flux. Indeed,
following the logic of \cite{lmg} and using the IIA vacua from section 4, 
it is possible to construct such an object.

We would like to take the conjecture one step further and motivated by the
fact that the superpotential can be written in a unified fashion for IIA
and IIB using pure spinors, introduce the ``pure spinor superfield", which
for IIA  will have $e^{iJ}$ as its lowest component and for IIB
reduces to the three-form superfield of \cite{lmg}:
$$
\varphi_1 (\theta, {\tilde \theta}) = \varphi_1 + \theta^2 (d + H
\bullet)\varphi_2 + {\tilde
\theta}^2 (d-H\bullet)\varphi_2 + \theta {\tilde \theta}
(e^{\phi}d(e^{-\phi}\varphi_2 + iC) \varphi_1) +...
$$
where as usually for IIA  $ \varphi_1 = {\rm exp }(iJ)$ and $ \varphi_2 =
\Omega$  and the inverse for IIB, and $C$ is the total RR field. The
$\theta\tilde\theta$ component can be obviously changed by
linear redefinitions such as $\theta \to \theta + \tilde\theta$ to get
an alternative form $(H\bullet + C) \varphi_2$, which agrees with the
three--form superfield above. 


\bigskip
\bigskip

{\bf Acknowledgement} We would like to thank Andrew Frey, Jan Louis, Simon Salamon,
Maxim Zabzine and Alberto Zaffaroni for useful discussions.

This work is supported in part by EU contract
HPRN-CT-2000-00122 and by INTAS contracts 55-1-590 and 00-0334.
MG was partially supported by European Commission Marie Curie Postdoctoral
Fellowship under contract number MEIF-CT-2003-501485.


\bigskip
\bigskip
\bigskip

\noindent
{\Large{\bf Appendix: $\,\,\,$ Derivation of the supersymmetry equations
in terms 
of Clifford products}}

\appendix
\renewcommand{\theequation}{A.\arabic{equation}}
\setcounter{equation}{0}\setcounter{section}{0}

\bigskip


\noindent
In this section we show how to derive expressions 
(\ref{eq:RSIIA})-(\ref{eq:RSIIB}) 
for the tensors $S,S_m,R_m$ and $R_{mn}$ introduced in (\ref{eq:susyRQ}).
We will perform  the derivation for IIA; the case of IIB works
analogously.

To treat the supersymmetry constraints for IIA and IIB in the most symmetrical 
way, it is convenient to consider linear combinations of the equations 
(\ref{eq:susyg}), (\ref{eq:moddil}) for the two spinors $\epsilon_1$ and
$\epsilon_2$. 
Inserting the Ansaetze (\ref{eq:metr}), (\ref{IIAans}), (\ref{eq:basissp}) and 
(\ref{eq:rrfs}) for the metric, the spinors and the RR field strengths, 
the equation for the space-time components of the gravitino, 
$\delta \psi_{\mu}$, can be rewritten as
\bea \label{IIAspf} 
\alpha \sla {\partial}A \eta_+  
+ \frac{i}{4} e^{\phi} \sla \!{F}_{A1}\eta_- &=& 0 \, ,\nn \\ 
\beta \sla {\partial}A \eta_+  
+ \frac{i}{4} e^{\phi}  \sla \!{F}_{A2} \eta_- &=& 0 \, ,
\eea 
where the coefficients $\alpha$ and $\beta$ are related to those in
(\ref{eq:basissp}) by 
\beq
\alpha \equiv a+ib \ \ \ \beta \equiv a-ib \, ,
\eeq
and $F_{A1}$ and $F_{A2}$ are those defined in (\ref{defGA}).
 
Repeating the procedure for the internal gravitino, 
$\delta \psi_m=0$, we obtain  
\beq \label{IIAintf} 
\alpha D_m(\eta_+) + ( \partial_m \alpha 
+ \frac{1}{4} \beta \sla \! {H}_m ) \eta_+ 
+ \frac{i}{8} e^{\phi} \sla \!{F}_{A1} \gamma_m  \eta_- = 0 \, ,
\eeq 
and the same expression with $\alpha \leftrightarrow \beta$ 
and $F_{A1} \leftrightarrow  F_{A2}$.

Similarly, for the modified dilatino equation one has 
\beq \label{IIAdilf} 
\alpha \sla \! D (\eta_+) 
+ \Big[\alpha \sla {\partial} (2A - \phi + log \alpha) 
+\frac{1}{4} \beta \sla \! H \Big] \eta_+ =0 \, ,
\eeq 
and again the same thing with $\alpha \leftrightarrow \beta$. 

For completeness we also list  the corresponding equations for IIB: 
\beq \label{IIBspf} 
\left[\alpha \sla {\partial} A + \frac{i}{4} e^{\phi} \sla
  \!{F}_{B1} \right] \eta_+ = 0  \, ,
\eeq 
\beq\label{IIBintf} 
\alpha D_m(\eta_+) + \left[ \partial_m \alpha -\frac{1}{4} \beta H_m -
  \frac{i}{8} e^{\phi} 
\sla \!{F}_{B1} \gamma_m \right] \eta_+ = 0  \, ,
\eeq 
\beq \label{IIBdilf}
\alpha \sla \! D (\eta_+) + \Big[\alpha \sla {\partial} (2A - \phi + log
\alpha) - \frac{1}{4} \beta \sla \! H \Big] \eta_+ =0 \, ,
\eeq 
where $\alpha$ and $\beta$ are defined as for IIA, $F_{B1}$ is defined 
in (\ref{defGB}), and we have again a second set of
equations with $\alpha \leftrightarrow \beta$ and 
$F_{B1} \leftrightarrow F_{B2}$ .
 
We want to write these equations in terms of Clifford products of the 
NS/RR fluxes with the two pure spinors $e^{iJ}$ and $\Omega$. 
As mentioned in the introduction, using Fierz rearrangement 
\beq 
\eta_{\pm} \otimes \eta^{\dagger}_+ = 
\frac{1}{4} \sum_{k=0}^6 \frac{1}{k!}  
\eta^{\dagger}_+ \gamma_{i_1 ... i_k} \eta_{\pm} 
\gamma^{i_k ... i_1} 
\label{eq:fz}
\eeq  
it is possible
to express the pure spinors as tensor products of the standard spinor 
defining the SU(3) structure
\bea \label{Fierz} 
\eta_{\pm} \otimes \eta^{\dagger}_{\pm} &=& 
\frac{1}{8} 
e^{\mp i J}\!\!\!\!\! 
\begin{picture}(10,10)
\put(0,0){\line(1,2){5}}
\end{picture} \, ,\nn \\ 
\eta_{+} \otimes \eta^{\dagger}_{-} &=& 
-\frac{i}{8} \sla {\Omega} \, ,\nn \\ 
\eta_{-} \otimes \eta^{\dagger}_{+} &=& 
-\frac{i}{8} \sla \overline{\Omega} \, .
\eea 

We can then rewrite equations (\ref{IIAspf}), (\ref{IIAintf}) 
and (\ref{IIAdilf}) in terms of Clifford products by multiplying them
to the left by $\eta^{\dagger}_{\pm}$ and $\eta^{\dagger}_{\pm} \gamma_n$. 
Take for example the spacetime gravitino
variation (\ref{IIAspf}), which we want to rewrite in the 
form (\ref{eq:susyRQ}), i.e. 
\beq \label{defS} 
\alpha \sla {\partial}A \eta_+  
- \frac{i}{4} e^{\phi} \sla \!{F}_{A1}\eta_- 
= S_{A1} \eta_- + (S_{A1\,m} + A_m) \gamma^m \eta_+ \, .
\eeq 
To obtain $S_{A1}$, 
we multiply this equality  
by $\eta_-^{\dagger}$ to the left on both sides. On the right hand side 
only $\frac{1}{2} S_{A1}$ remains, while on the left hand side the first 
term goes away, and the other can be written as
\beq 
\eta_-^{\dagger} \sla \!F_{A1} \eta_- = Tr (\eta_-^{\dagger}\sla \! F_{A1} 
\eta_-)= \frac{1}{8}Tr(\sla \!F_{A1} \sei)=  
\frac{1}{2}(\sla \!F_{A1} \sei)_0 \, .
\eeq   
In the first equality we inserted a trace since the lhs is a number, while
in the second one 
we have used the cyclic property of the trace and (\ref{Fierz}).
Finally in the last step we used the fact 
that antisymmetric products of gamma matrices are traceless, 
and that the gammas in six dimensions are 4 x 4 matrices, so $Tr {\bf 1}=4$.  
By $(...)_0$ we mean the term in the Clifford product 
that does not contain any gamma matrix.  
 
Similarly, $S_{A1\,m}$ and $A_m$ are obtained multiplying (\ref{defS}) 
by $\eta_+^{\dagger} \gamma_p$. 
On the right hand side we have $\overline{P}_p \,^m S_m$,  
where 
$\overline{P}$ is the projector onto antiholomorphic indices 
$\overline{P}_n\,^m= \frac{1}{2} (\delta_n\,^m + i J_n\,^m)$, 
while on the left hand side we get the warp 
factor term 
and a contribution from the RR fluxes. To evaluate this contribution, we need 
the identity  
\beq 
\eta_+^{\dagger} \gamma_p \sla \!F_A \eta_- =  
Tr (\eta_+^{\dagger} \gamma_p \sla \! {F}_A  \eta_-)=  
-\frac{i}{8}Tr(\sla \overline{\Omega}  \gamma_p \sla \! {F}_A)= 
- \frac{i}{2}  (\sla \! {F}_A \sla \overline{\Omega})_p \, ,
\eeq 
where $(...)_p$ means the term that multiplies $\gamma^p$ in the 
Clifford product. 
 
Collecting everything together, the spacetime gravitino variations 
(\ref{IIAspf}) can be written as 
\bea 
\frac{i}{4} e^{\phi} (\sla \!F_{A1} \sei)_0 \eta_- 
+ \Big[\frac{1}{8} e^{\phi} 
 {\rm Re}(\sla \!F_{A1}  \sla  \overline{\Omega})_m 
+ \alpha \partial_m A \Big]\gamma^m \eta_+ &=& 0 \, ,
\eea  
plus the same equation with $F_{A1} \rightarrow F_{A2}$ and $\alpha \to \beta$.
From this equation we can immediately read  $S$, $S_{m}$ and $A_m$ as
given in \ref{eq:RSIIA}. 

 
As for the internal gravitino equations (\ref{IIAintf}), we want to write 
them in the form (\ref{eq:susyRQ})
\beq \label{defR} 
Q_m \eta_+ + Q_{mp} \gamma^p \eta_- + R_m \eta_+ + R_{mp} \gamma^p \eta_- 
\eeq 
where $Q_m$ and $Q_{mp}$ summarize the torsions and NS-flux contribution. 
Their expressions can be found in (\ref{eq:RSIIA}).

To get $R_m$ and $R_{mp}$ we should  multiply (\ref{IIAintf}) 
by $\eta^{\dagger}_+$ and  
$\eta^{\dagger}_- \gamma_n$ on the left, respectively. Doing this on
(\ref{defR}), we obtain  $\frac{1}{2} R_m$ 
and $P_n\,^p R_{mp}$, where $P$ is the  
projector onto holomorphic indices $P_n\,^m= \frac{1}{2} (\delta_n\,^m - i
J_n\,^m)$.
 
We repeat this procedure for all supersymmetry equations which can now be
written in a nice and compact way
\beq 
\label{IIAspclif}
\frac{i}{4} e^{\phi} (\sla \!F_{A1} \sei)_0 \, \eta_- 
+ \Big[\frac{1}{8} e^{\phi}  
(\sla \!F_{A1}  \sla  \overline{\Omega})_m  
+ \alpha \partial_m A \Big]\gamma^m \eta_+ = 0 \, ,
\eeq
\bea \label{IIAintclif}
\frac{i}{8} e^{\phi} 
\left[\frac{1}{2}(\sla \!F_{A1} \sei)_0 g_{mp} 
+ (\sla \!F_{A1} \sei)_{mp}
- (  \sla \!F_{A1 \, m} \sei)_p \right] \gamma^p \eta_- +\nn\\
+ i{ Q}_{1\,m} \eta_+ + i{ Q}_{1\,mp} \gamma^p \eta_- + \frac{1}{8} e^{\phi}  
( \sla  \overline{\Omega} \sla \!F_{A1})_m \eta_+  =0  \, ,
\eea 
\bea \label{IIAdilclif}
\Big[ i \alpha { q}_m + \frac{i}{2} \alpha q_{nr}  \Omega^{nr}\,_m 
+ \frac{1}{48}\beta 
 (\sla \! H \seim )_m 
+ \alpha {\partial}_m (2A - \phi+ ln \alpha)\Big] \gamma^m \eta_+  + \nn\\
+ \Big[2i \alpha P^{mn} q_{mn} 
-\frac{i}{24}\beta (\sla \! H \Omega)_0 \Big] \eta_- =0 
\eea  
for IIA, and for IIB
 \bea \label{IIBspclif}
\frac{1}{4} e^{\phi}  (\sla \!F_{B1} \sla \Omega)_0 \, \eta_- 
+ \Big[\frac{i}{8} e^{\phi}(\sla \!F_{B1} \seim)_m  
+ \alpha \partial_m A \Big]\gamma^m \eta_+ = 0  \, ,
\eea 
\bea \label{IIBintclif}
i{ Q}_{1\,m}\, \eta_+ + i{ Q}_{1\, mp} \,\gamma^p \eta_- 
- \frac{i}{8} e^{\phi}  
( \seim \sla \!F_{B1})_m \eta_+  \ \ \ \ \ \nn\\ 
+ \frac{1}{8} e^{\phi} \left[- (  \sla \!F_{B1 \, m} \sla \Omega)_p  + 
\frac{1}{2}(\sla \!F_{B1} \sla \Omega)_0 g_{mp} 
+ (\sla \!F_{B1} \sla \Omega)_{mp} \right] \gamma^p \eta_- =0 \, ,
\eea
\bea \label{IIBdilclif}
\Big[2i \alpha P^{mn} q_{mn}  
+\frac{i}{24}\beta (\sla \! H \Omega)_0 \Big] \eta_-  \ \ \ \ \ \nn\\
+\Big[i \alpha { q}_m + \frac{i}{2} \alpha q_{nr} 
\overline{\Omega}^{nr}\,_m - \frac{1}{48}\beta 
 (\sla \! H e^{-i \sla J} )_m 
+ \alpha {\partial}_m (2A - \phi+ln \alpha)\Big] \gamma^m \eta_+ =0 \, .
\eea
In deriving the  equations above we used the following identities:
\bea 
\eta_-^{\dagger} \sla \!F_A \eta_- &=& 
\frac{1}{2}(\sla \!F_A \sei)_0 \, , \\ 
\eta_+^{\dagger} \gamma_m \sla \!F_A \eta_- &=&  
- \frac{i}{2}  (\sla \! {F}_A \sla \overline{\Omega})_m \, , \nn \\ 
 \eta_+^{\dagger} \sla \!{F}_A \gamma_m \eta_- &=& 
-\frac{i}{2}  (\sla \overline{\Omega} \sla \! {F}_A )_m  \, , \nn\\  
 \eta_-^{\dagger} \gamma_p \sla \!{F}_A \gamma_m \eta_- &=&  
- (  \sla \! {F_{A\,m}} \sei )_p 
+ \frac{1}{2} ( \sla \! {F_A}  \sei)_0 \, g_{mp} 
+ (\sla \! {F_A} \sei)_{mp} \nn
\eea 
for IIA, and for IIB 
\bea 
\eta_-^{\dagger} \sla \!F_B \eta_+ &=& 
-\frac{i}{2}(\sla \!F_B \sla {\Omega} )_0 \, , \\ 
\eta_+^{\dagger} \gamma_m \sla \!F_B \eta_+ &=&   
\frac{1}{2}  (\sla \! {F}_B \seim)_m \, , \nn \\ 
 \eta_+^{\dagger} \sla \!{F}_B \gamma_m \eta_+ &=& 
\frac{1}{2}  (\seim\sla \! {F_B})_{m} \, ,  \nn\\ 
 \eta_-^{\dagger} \gamma_p \sla \!{F}_B \gamma_m \eta_+ &=&   
-i (   \sla \!  {F_{B\,m}} \sla \! \Omega )_p 
+ \frac{i}{2} ( \sla \! {F}_B   \sla \! \Omega)_0 \, g_{mp}  
+ i(\sla \! {F}_B  \sla \! {\Omega})_{mp} \nn \, .
\eea 

The explicit expressions for the Clifford products appearing 
in (\ref{IIAspclif})-(\ref{IIBdilclif}) are:
\bea \label{clif1}
(\sla \!F_{A} \sei)_0 &=& F_0 
- \frac{i}{2} F^{ab}J_{ab} - \frac{1}{8} F^{abcd} J_{ab} J_{cd} 
+ \frac{i}{48} F^{abcdef}J_{ab}J_{cd}J_{ef} \, , \nn \\
(\sla \!F_{A}  \sla  \overline{\Omega})_m &=&
- \frac{1}{2} F^{ab} \overline{\Ox}_{abm} 
+ \frac{1}{6} F^{abc}\,_m \overline{\Ox}_{abc} \, , \nn \\
( \sla  \overline{\Omega} \sla \!F_{A })_m &=& 
-\frac{1}{2} F^{ab} \overline{\Ox}_{abm} 
-\frac{1}{6} F^{abc}\,_m \overline{\Ox}_{abc}  \, , \nn \\
( \sla \!F_{A \, m} \sei)_n &=& 
2 F_{mp} P_n\,^p + i J_{ab} F^{ab}\,_{mp} \overline{P}_n\,^p
 - \frac{1}{4} J_{ab}J_{cd} F^{abcd}\,_{mp} P_n\,^p \, ,  \nn\\
(\sla \!F_{A} \sei)_{mn} &=& \frac{i}{2} F_0 J_{mn} 
+\frac{1}{2} F_{mn} +i F^a\,_{[m} J_{n]a} + \frac{1}{4} F^{ab}J_{ab} J_{mn} 
- \frac{1}{2} F^{ab} J_{am} J_{bn} \nn \\
&&   - \frac{i}{4} F_{mn}\,^{ab} J_{ab} 
- \frac{1}{2} F^{abc}\,_{[m} J_{n]a} J_{bc}
-\frac{i}{16} F^{abcd} J_{ab}J_{cd}J_{mn}
+ \frac{i}{4} F^{abcd} J_{ab}J_{cm}J_{dn} \nn\\
&& + \frac{1}{16} F^{abcd}\,_{mn} J_{ab} J_{cd} 
\eea
for IIA, where $F_A$ is a generic sum of fluxes, i.e. $F_A= F_{(0)} + F_{(2)} + F_{(4)}+ F_{(6)}$, and 
\bea \label{clif2}
(\sla \!F_{B} \sla  {\Omega})_0 &=& -\frac{1}{6} F^{abc} \Ox_{abc} \, ,\nn \\
(\sla \!F_{B} \seim )_m &=& 2 F_n  \overline{P}_m\,^n + i J_{ab} F^{ab}\,_{n} \overline{P}_m\,^n 
- \frac{1}{4} J_{ab} J_{cd} F^{abcd}\,_n  \overline{P}_m\,^n  \, , \nn \\
( \seim\sla \!F_{B })_m &=&  2 F_n  {P}_m\,^n + i J_{ab} F^{ab}\,_{n} {P}_m\,^n 
- \frac{1}{4} J_{ab} J_{cd} F^{abcd}\,_n  {P}_m\,^n  \, , \nn \\
(  \sla \!F_{B \, m}\sla  {\Omega})_n &=& -\frac{1}{2} F^{ab}\,_m \,\Ox_{nab} 
-\frac{1}{6} F^{abc}\,_{mn} \Ox_{abc} \, , \nn\\
(\sla \!F_{B} \sla  {\Omega})_{mn} &=& \frac{1}{2} F^{a} \Ox_{amn} -
\frac{1}{2} F^{ab}\,_{[m} \Ox_{n]ab}- \frac{1}{12} F^{abc}\,_{mn} \Ox_{abc} 
\eea
for IIB, where $F_B= F_{(1)} + F_{(3)} + F_{(5)}$. Finally
\bea \label{clifH}
(\sla \!H \sla  {\Omega})_0 &=& -\frac{1}{6} H^{abc} \Ox_{abc} \, , \nn \\
(\sla \!H \seim )_m &=& i J_{ab} H_n\,^{ab} \overline{P}_m\,^n  
\eea
for both IIA and IIB. 

In the text we have used the decomposition of the fluxes 
in SU(3) representations. These are defined in the 
following way: for the $H$-flux 
\beq
 H = -\frac{3}{2}\, {\rm Im}(H^{(1)} \bar{\Omega}) + H^{(3)} \wedge J 
+ H^{(6)} \, ,
\end{equation} 
and similarly for the IIB RR 3-form flux. The components are explicitly given by
\bea
H^{(1)}&=& -\frac{i}{36}   H^{ijk} \Ox_{ijk} \, ,\nn \\
H^{(3)}_i &=& \frac14 H_{imn} J^{mn} \, , \nn \\
H_{ij}^{(6)} &=& H^{kl}\,_{(i} \Ox_{j)kl} \, .
\eea
RR three-form flux $F_3$ decomposes exactly in the same way as $H$. For
the rest of the RR fluxes we use the following decompositions
\bea
F_{2}&=& \frac{1}{3} F_{2}^{(1)}\, J +  Re(F_{2}^{(3)}\llcorner
\overline{\Ox}) + F_{2}^{(8)}  \, , \nn \\
F_{4}&=& \frac16 F_{4}^{(1)} J \wedge J +   Re(F_{4}^{(3)}\wedge
\overline{\Ox}) + F_{4}^{(8)} \, , \nn\\
F_{5} &=&  F_{5}^{(3)} \wedge J \wedge J \, , \nn\\
F_{6} &=& \frac{1}{6}  F_{6}^{(1)} \,J \wedge J \wedge J \, ,
\eea
where the different representations are given by
\bea
F_{2}^{(1)}&=& \frac{1}{2} F_{mn} J^{mn} =
F_{i \bar{j}} J^{i \bar{j}} \, , \nn \\
F_{2\,k}^{(3)} &=& \frac{1}{8} F^{ij} \Ox_{ijk} \, , \nn\\
F_{4}^{(1)}&=&\frac18 F^{mnpq} J_{mn} J_{pq} \, , \nn \\
F_{4\,k}^{(3)}&=& \frac{1}{24} F_{k}\,^{ijl} \Ox_{ijl} \, , \nn \\
F_{5\,i}^{(3)}&=& \frac{1}{16} F_i^{mnpq} J_{mn} J_{pq} \, ,   \nn\\
 F_{6}^{(1)}&=& \frac{1}{48} F^{mnpqrs} J_{mn} J_{np} J_{qr} \, .
\eea

With the definitions above, we can write the matrices $Q$, $R$ and $S$ 
in terms of SU(3) representations.
For IIA we have
\bea \label{eq:RSIIAb}
A_{\bi} &=& \alpha \partial_{\bi} A \, , \nn \\
S&=&- \frac{i}{4} e^{\phi} \left[\beta^* F_0
- i \alpha^* F^{(1)}_2
- \beta^*  F^{(1)}_4
+ i \alpha^*  F^{(1)}_6 \right] \, ,
\nn \\
%
%
S_{\bi}&=&\frac{1}{2} e^{\phi}
(\alpha^* F^{(\bar{3})}_2 +
\beta^* F^{(\bar{3})}_4)_{\bar{i}} \, ,  \nn \\
Q_{\bar i}&=& -i \partial_{\bi} \alpha -\frac{i}{2} \Big[ \alpha (W_5 - W_4) +
i \beta H^{(3)} \Big]_{\bi}  \, , \nn \\
Q_{ij}&=& - \frac{1}{8} \Big[ \Omega_{ijk} (\alpha W_4 +i \beta H^{(3)})^k
+ \frac{i}{2} (\alpha W_3
+ i \beta H^{(6)})_{i}^{\, \, \, kl} \Omega_{jkl} \Big] \, , \nn\\
Q_{\bi j}&=& \frac{1}{4} \Big[(\alpha W_1 - 3 i \beta H^{(1)}) g_{\bi
  j} + i \alpha W_{2\,\bi j} \Big]\, ,  \nn\\
R_{i} &=& 0 \, , \nn\\
R_{\bar{i}}&=&  \frac{i}{2} e^{\phi}
(\alpha^* F^{(\bar{3})}_2 -
\beta^* F^{(\bar{3})}_4)_{\bar{i}} \, , \nn \\
R_{ij}&=& 0 \, , \nn \\
R_{\bi j}&=& - \frac{1}{8} e^{\phi} \left[ g_{\bar{i}j} \bar S - g_{\bar{i}j} 
(\frac{8}{3} \beta F^{(1)}_4 - \frac{8}{3} \alpha^*  F^{(1)}_6)
-2 \alpha^* F^{(8)}_{2\, \bi j}  
-2 i \beta F^{(8)}_{\bi j \bar k l} J^{\bar k l}
\right] \, , 
 \nn\\
T&=& \frac{3}{4} (\alpha i W_1 - \beta H^{(1)}) \, , \\
T_{\bi} &=&  \alpha \partial_{\bi} (2A -\phi - \log \alpha)
+ \frac{1}{2} \Big[ \alpha (W_4 + W_5) -i \beta H^{(3)} \Big]_{\bi} \, .
\eea

For IIB we get
\bea \label{eq:RSIIBb}
A_{\bi}&=& \alpha \partial_{\bi} A \, , \nn \\
S&=&\frac{3}{2} i \beta e^{\phi} F_3^{(1)} \, , \nn \\
S_{i}&=& \frac{1}{4} e^{\phi}
\Big(\alpha^* F^{(3)}_1 + 2 i \beta^* F_3^{(3)} 
- 2 \alpha^* F_5^{(3)} \Big)_i \, , \nn \\
S_{\bi}&=& \frac{1}{4} e^{\phi}
\Big(\alpha F^{(\bar{3})}_1 - 2 i \beta F_3^{(\bar{3})}
- 2 \alpha F_5^{(\bar{3})} \Big)_{\bi} \, , \nn \\
Q_{\bar i}&=& -i \partial_{\bi} \alpha -\frac{i}{2} \Big[ \alpha (W_5 - W_4) -
i \beta H^{(3)} \Big]_{\bi} \, ,  \nn \\
Q_{\bi j}&=& \frac{1}{4} \Big[(\alpha W_1 + 3 i \beta H^{(1)}) g_{\bi
  j} + i \alpha W_{2\,\bi j} \Big] \, , \nn\\
Q_{ij}&=&  
- \frac{1}{8} \Big[ \Omega_{ijk} (\alpha W_4 -i \beta H^{(\bar 3)})^k
+ \frac{i}{2} (\alpha W_3
- i \beta H^{(6)})_{i}^{\, \, \, kl} \Omega_{jkl} \Big] \, , \nn\\
R_i&=& - \frac{i}{4} e^{\phi}
\Big(\alpha F^{(3)}_1 -  2 i \beta F_3^{(3)} 
- 2 \alpha  F_5^{(3)} \Big)_i  \, , \nn \\
R_{\bi}&=& 0 \, , \nn \\
R_{ij}&=& - \frac{i}{16} e^{\phi} \Big(
\alpha F^{(\bar{3})k }_1 \Omega_{ijk}
- \beta F^{(6)}_{3 ij}
+2 \alpha F_5^{(\bar{3}) k}\Omega_{ijk} \Big) \, , \nn\\
R_{\bi j}&=& 0 \, , \nn\\
T&=& \frac{3}{4} (\alpha i W_1 + \beta H^{(1)}) \, , \\
T_{\bi} &=&  \alpha \partial_{\bi} (2A -\phi - \log \alpha)
+ \frac{1}{2} \Big[ \alpha (W_4 + W_5) +i \beta H^{(3)} \Big]_{\bi} \, .
\eea


\end{document}